\documentclass[a4paper,10pt]{article}
\usepackage[utf8]{inputenc}
\usepackage{graphics}
\usepackage{amsmath}
\usepackage{enumerate}
\usepackage{siunitx}
\usepackage{stmaryrd} 
\usepackage{dsfont} 
\usepackage{graphicx}
\usepackage{enumerate}
\usepackage{subfigure}
\usepackage{doi}
\usepackage{cite}

\usepackage{mdframed}
\usepackage{placeins}

\usepackage{acronym}
\usepackage{mhchem}

\DeclareMathOperator{\mum}{\mu \text{m}} 

\newcommand{\ddt}[1]{\frac{d #1}{dt}}

\newcommand{\RR}{\mathds{R}} 

\newcommand{\visc}{\eta} 
\DeclareMathOperator{\sigmawet}{\sigma_w}

\DeclareMathOperator{\thetaeq}{\theta_{eq}}
\DeclareMathOperator{\thetaapp}{\theta_{app}}
\DeclareMathOperator{\Ca}{Ca} 
\DeclareMathOperator{\eqlength}{L_{eq}} 
\DeclareMathOperator{\clspeed}{U_{cl}} 

\usepackage{amsthm}
\usepackage[a4paper,left=2.5cm,right=5cm,top=3.5cm,bottom=5cm]{geometry}
\usepackage{authblk} 

\usepackage[margin=1cm]{caption} 

\date{}

\theoremstyle{definition}
\newtheorem{example}{Example}
\newtheorem{remark}{Remark}

\title{A geometry-based model for spreading drops\\ applied to drops on a silicon wafer and a swellable polymer brush film}

\author[1]{M. Fricke\thanks{fricke@mma.tu-darmstadt.de}}
\author[2]{B. Fickel}
\author[3]{M. Hartmann}
\author[1]{D. Gründing}
\author[2]{M. Biesalski} 
\author[1]{D. Bothe}

\affil[1]{Mathematical Modeling and Analysis Group, TU Darmstadt}
\affil[2]{Macromolecular and Paper Chemistry, TU Darmstadt}
\affil[3]{Nano- and Microfluidics, TU Darmstadt}

\begin{document}

\maketitle

\abstract{
We investigate the dynamics of spreading in a regime where the shape of the drop is close to a spherical cap. The latter simplification is applicable in the late (viscous) stage of spreading for highly viscous drops with a diameter below the capillary length. Moreover, it applies to the spreading of a drop on a swellable polymer brush, where the complex interaction with the substrate leads to a very slow spreading dynamics. The spherical cap geometry allows to derive a closed ordinary differential equation (ODE) for the spreading if the capillary number is a function of the contact angle as it is the case for empirical contact angle models. The latter approach has been introduced by de Gennes (Reviews of Modern Physics, 1985) for small contact angles. In the present work, we generalize the method to arbitrary contact angles.\\
The method is applied to experimental data of spreading water-glycerol drops on a silicon wafer and spreading water drops on a PNIPAm coated silicon wafer. It is found that the ODE-model is able to describe the spreading kinetics in the case of partial wetting. Moreover, the model can predict the spreading dynamics of spherical cap-shaped droplets if the relationship between the contact angle and the capillary number is universal.
} 
%


\section{Introduction}
The spreading of droplets on solid surfaces is a prototypical situation in the field of dynamic wetting and has been studied extensively in the literature see \cite{Gennes.1985,Bonn.2009,Gennes.2010} and references therein for a review. The equilibrium state of a droplet on a surface depends strongly on the surface energy of the material which is closely related to the surface properties, chemistry and microstructure. By modifying the surface properties, one can therefore drastically change the wettability of the surface with a potentially large impact on technical processes. Printing technology and microfluidic devices are two prominent examples. Mathematically, the equilibrium state can be found by means of a minimization of the free energy functional consisting of a gravitational term and the interfacial energies of the liquid-gas, liquid-solid and solid-gas interfaces. It can be shown that for small values of the Bond number, defined as
\begin{equation} \label{eq:bond_number}
\text{Bo} = \frac{\rho \text{g} l^2}{\sigma_{\text{lg}}}
\end{equation}
with $\sigma_{\text{lg}} > 0$ the surface tension of the liquid-gas interface and $l$ a characteristic length scale, the stationary shape of the drop is close to a spherical cap. In case of a smooth and homogeneous substrate, the contact angle of the spherical cap satisfies the Young–Dupré equation \cite{Young.1805}, i.e.\
\begin{align}
\label{eqn:young_equation}
\sigma_{\text{lg}} \cos \theta + \sigmawet= 0,
\end{align}
where $\sigmawet = \sigma_{\text{ls}} - \sigma_{\text{sg}}$ is the specific energy of the wetted solid surface (relative to a dry surface). By introduction of the spreading coefficient $S$ according to \cite{Gennes.1985}, i.e.\
\begin{align*}
S := - (\sigmawet + \sigma_{\text{lg}}),
\end{align*}
one can equivalently rewrite Equation~\eqref{eqn:young_equation} as
\begin{align*}
\frac{S}{\sigma_{\text{lg}}} = \cos \theta - 1.
\end{align*}
The spreading parameter $S$ discriminates the cases of complete wetting characterized by $S>0$, where no solution to \eqref{eqn:young_equation} exists and the liquid spreads out completely, from the case of partial wetting characterized by $S < 0$, where a finite contact angle exists. Note that Equation~\eqref{eqn:young_equation} may be generalized to describe the stationary wetting in more general situations, for example on rough and chemically heterogeneous surfaces (known as Wenzel's Model \cite{Wenzel1936} and Cassie-Baxter Model \cite{Cassie1944}).\newline
\newline
An interesting wetting behavior occurs if the surface consists of a thin polymer film. To this end, surface wettability of covalently bonded graft polymer films, such as polymer brushes, can be tuned by the chemistry and swelling behavior of the attached macromolecules. Using surface initiated polymerization techniques, a large variety of chemical functions can be implemented into the polymer film, including those that are renowned as “smart” or “stimulus responsive” polymers. Significant research has been carried out to understand how molecular mass and grafting density affect the wetting and swelling in contact with a solvent \cite{doi:10.1002/adma.201706441, doi:10.1021/cr900045a}. Much less is known about the dynamics of wetting of such swellable polymer brushes, in particular, if a "pre-swelling" in humid air is considered. The latter impacts a possible imbibition of the solvent beyond the macroscopic contact line into the polymer brush, and forces chain relaxations (swelling) which occur on much different geometric and time scales, respectively. An interesting continuum mechanical model for the wetting on polymer brushes has been introduced recently in \cite{Thiele2019}.\\
\\
Even though the statics of wetting is well-understood for a long time (see \cite{Young.1805,Wenzel1936,Cassie1944}), the mathematical modeling of dynamic wetting is still a challenge. From the perspective of mathematical modeling, there is a whole hierarchy of models to describe dynamic wetting, ranging from molecular dynamics via continuum mechanics to simplified/empirical descriptions of the macroscopic flow. While the description on the molecular level can be expected to be the most accurate one in principle, it is limited for practical purposes due to the very high computational costs. Therefore, it is desirable to develop mathematical models that are able to describe the physics of wetting on larger scales. Typically, the latter is done in the framework of the Navier Stokes equations in a free-surface or multiphase formulation. While the continuum mechanical models have shown their ability to describe a large number of wetting flows quite accurately, they are still demanding in terms of computational costs and they produce a large amount of data that needs to be processed. A substantial amount of research is dedicated to finding \emph{empirical} correlations describing some macroscopic features of the flow with only a few parameters. The latter type of description might be useful to design systems for practical applications.

\begin{figure}[ht]
 \subfigure[Glycerol-water droplet on a silicon wafer.]{\includegraphics[width=6.5cm]{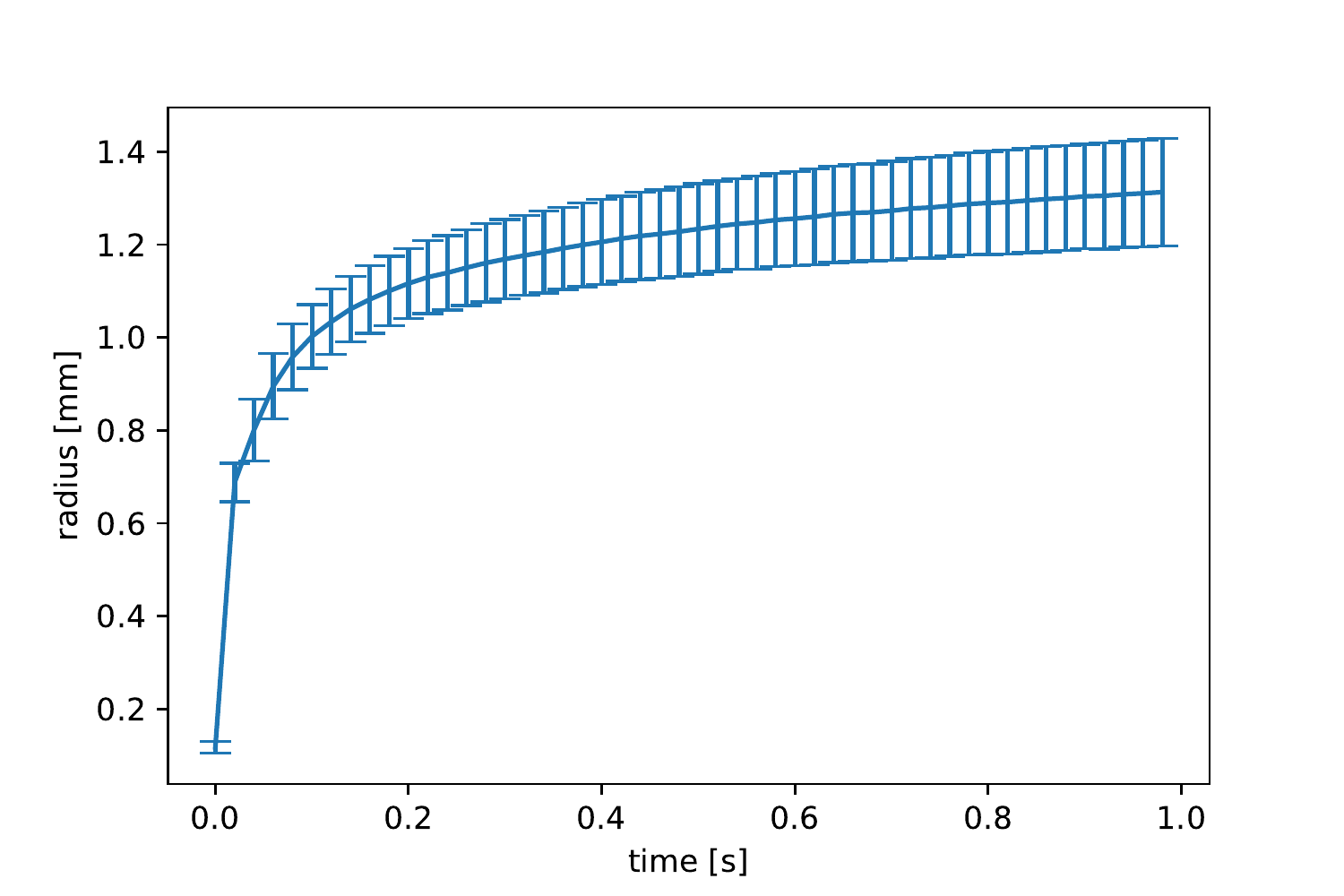}\label{fig:spreading_example_glycerol}}
 \subfigure[Water droplet on a PNIPAm polymer brush.]{\includegraphics[width=6.5cm]{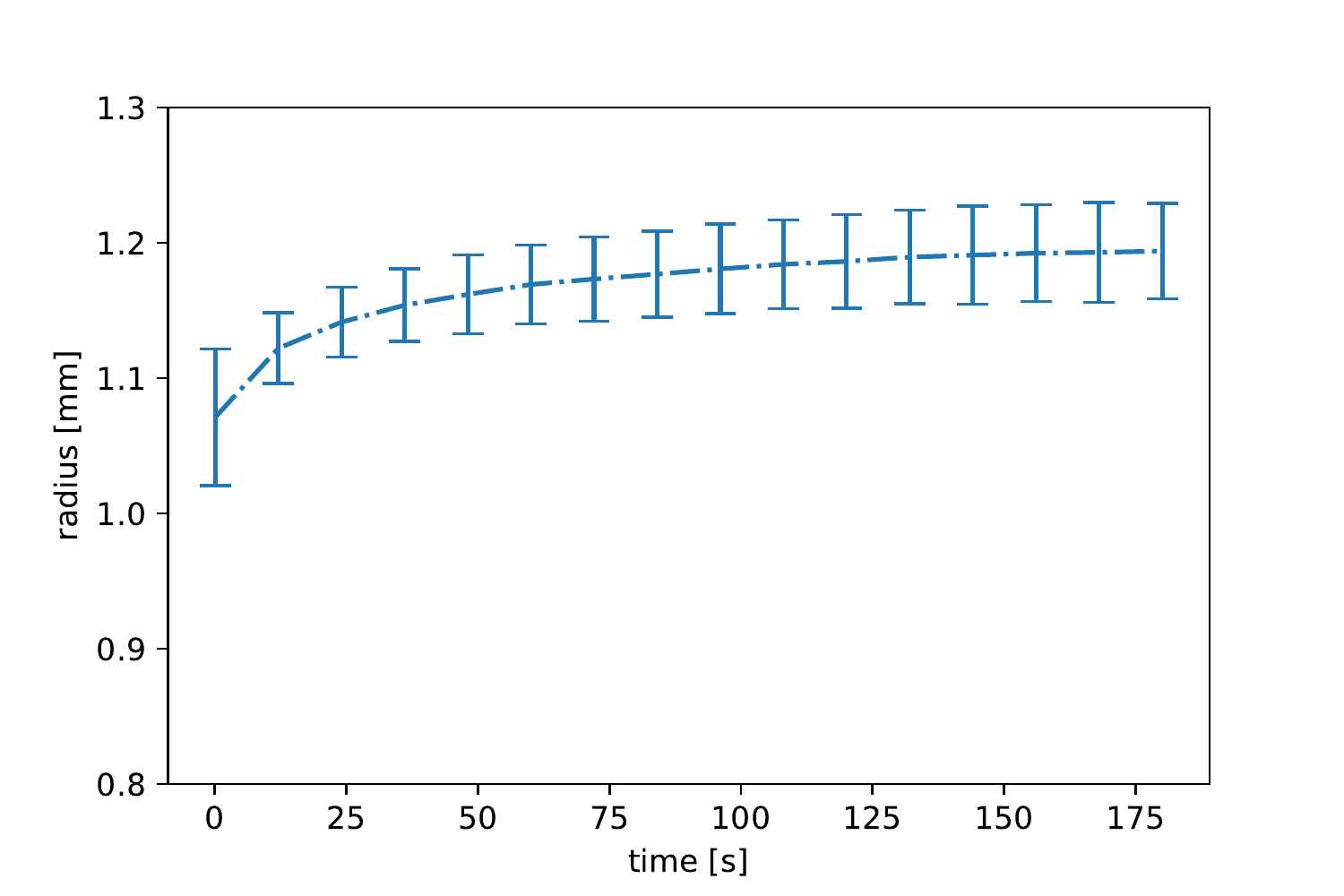}\label{fig:spreading_example_polymer-brush}}
 \caption{Experimental data for the spreading of a droplet.}
 \label{fig:radius_over_time_exp}
\end{figure}

The present work aims at an empirical mathematical model to describe the process of spreading in terms of the base radius of the droplet as a function of time. Figure~\ref{fig:radius_over_time_exp} shows experimental data for the spreading of a glycerol-water droplet on a silicon wafer and the spreading of a pure water droplet on a PNIPAm brush. Note that the timescale for the spreading dynamics on the polymer brush in Figure~\ref{fig:spreading_example_polymer-brush} differs in orders of magnitude from the timescale for the water-glycerol droplet in Figure~\ref{fig:spreading_example_glycerol}.

\paragraph{Assumptions:} The reasoning which has been formulated for the special case of flat drops in \cite{Gennes.1985} is based on the following assumptions:
\begin{enumerate}[(I)]
 \item The shape of the droplet throughout the spreading process is a spherical cap. 
 \item The speed of the contact line can be expressed as a function of the contact angle\footnote{Here, by ``contact angle'' we mean the ``macroscopic'' contact angle that corresponds to the spherical cap shape.} $\theta$ (or vice versa).
 \item The volume of the drop is known (as a function of time if it is not conserved).
\end{enumerate}
Note that the ordinary differential equation derived in Section~\ref{section:theory} is a direct consequence of (I)-(III) without further approximations or assumptions.

\begin{figure}[ht]
  \subfigure{\includegraphics[width=4cm]{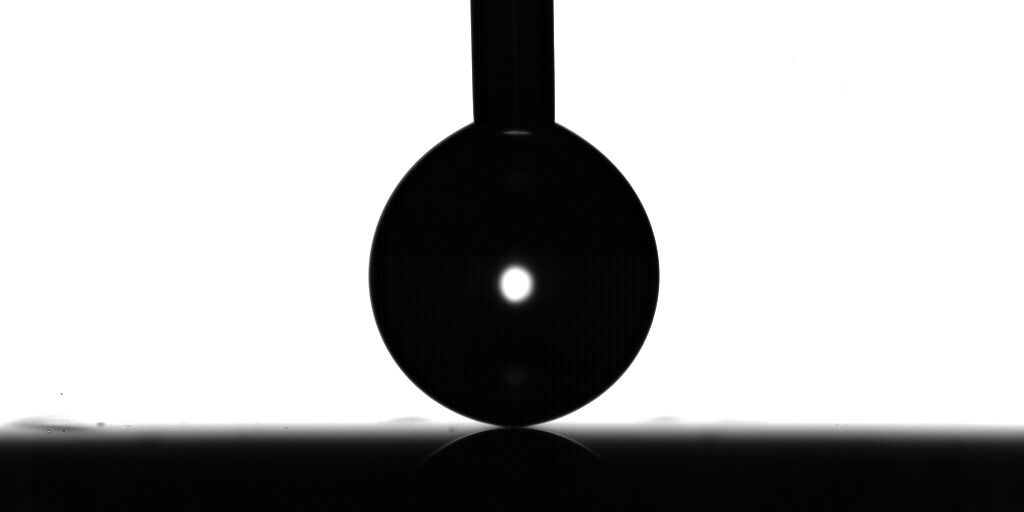}}
  \subfigure{\includegraphics[width=4cm]{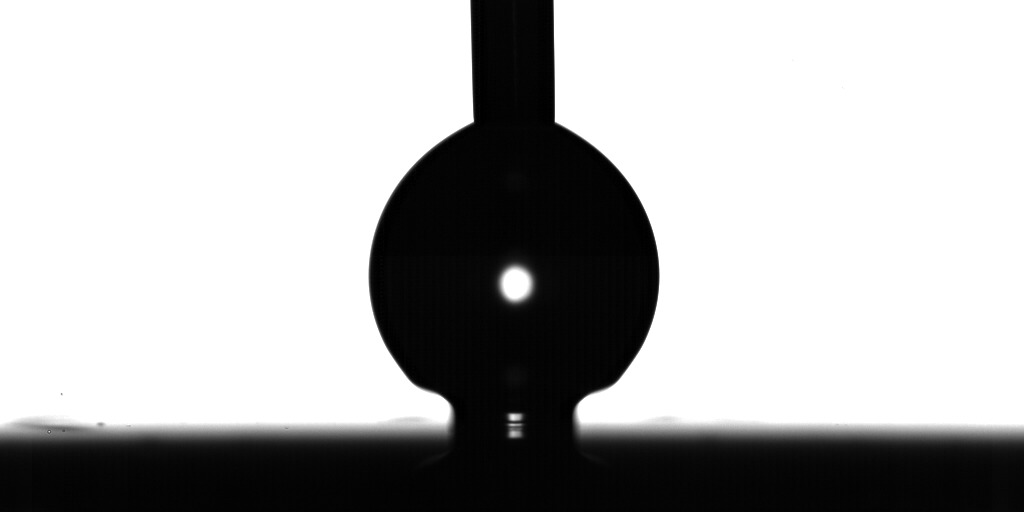}}
  \subfigure{\includegraphics[width=4cm]{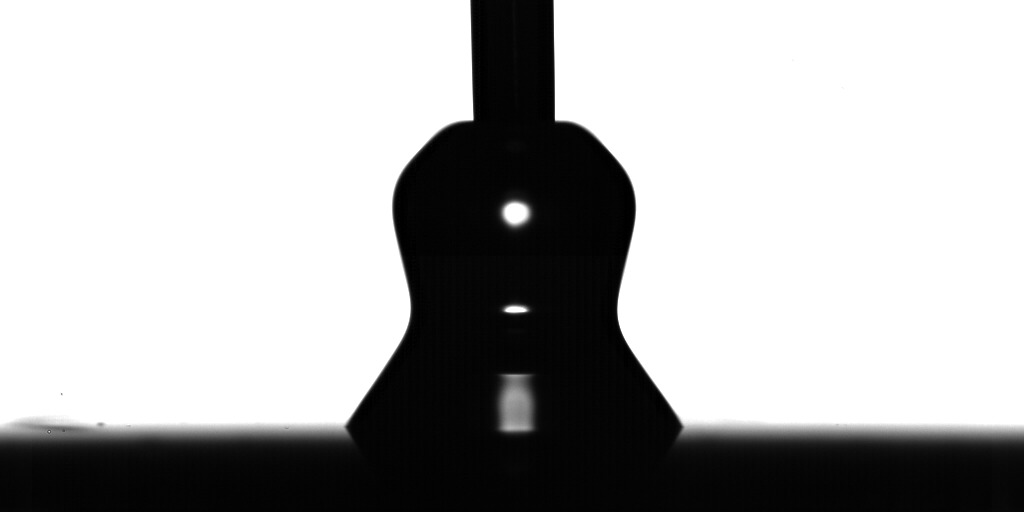}}\\
  \subfigure{\includegraphics[width=4cm]{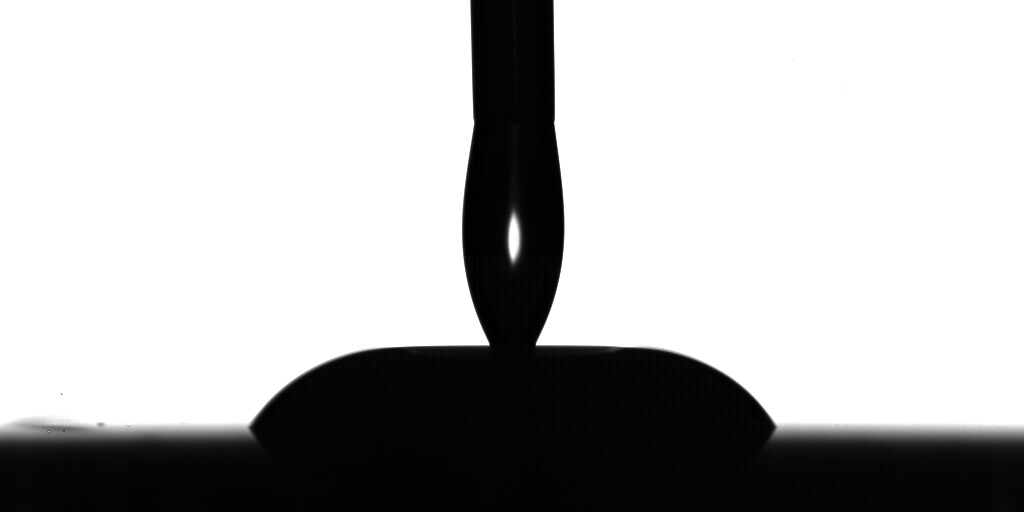}}
  \subfigure{\includegraphics[width=4cm]{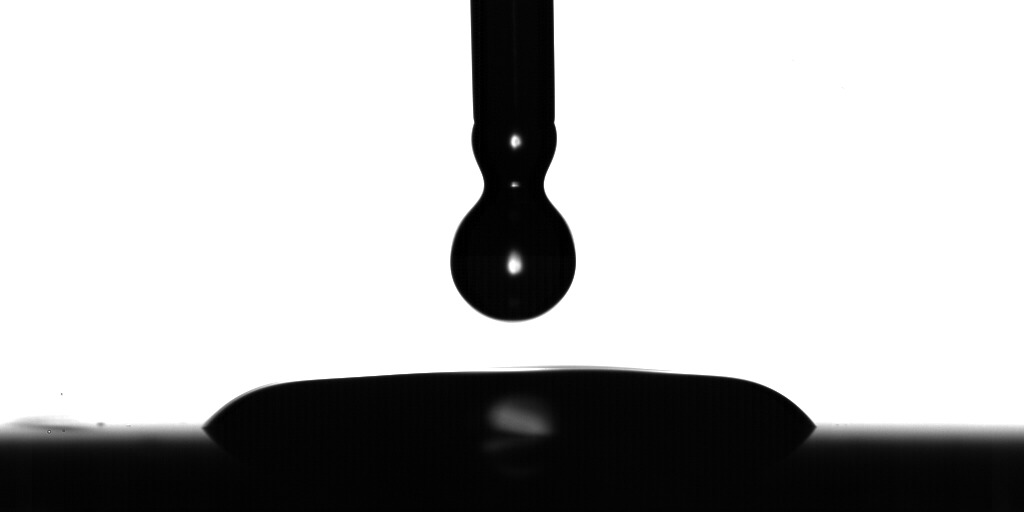}}
  \subfigure{\includegraphics[width=4cm]{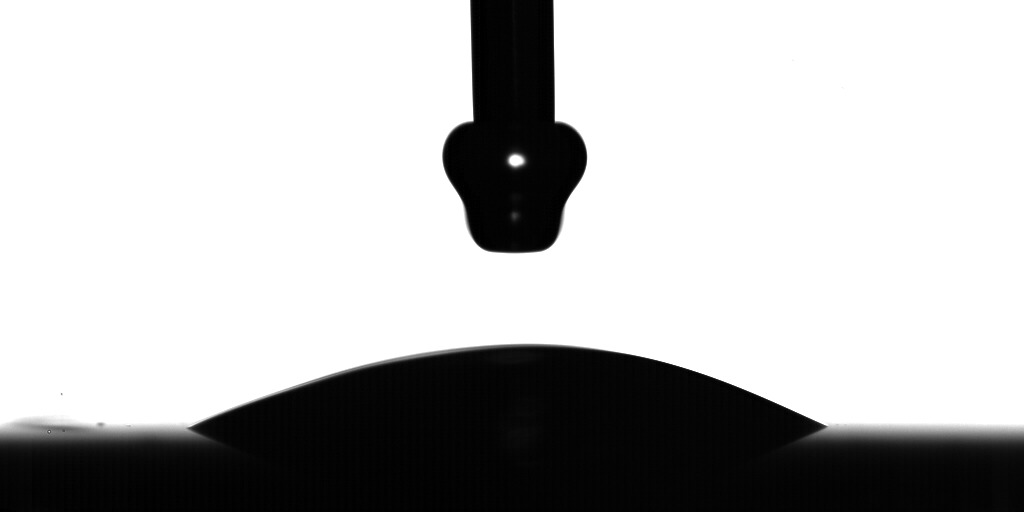}}
  \caption{Deposition of a pure water droplet on a silicon wafer.}
  \label{fig:deposit_water_droplet}
 \end{figure}

\paragraph{Remarks on assumption (I):} The first assumption (I) greatly simplifies the description since in this case, the shape is completely determined by only two parameters. Physically, the spherical cap is expected to be a good approximation in the late stage of spreading, i.e.\ for $\theta$ close enough to equilibrium if surface tension forces dominate over both gravity and viscous forces, i.e.\ if the Bond and capillary numbers defined as
\begin{align} 
\label{eqn:bo_and_ca}
\text{Bo} = \frac{\rho \text{g} l^2}{\sigma} = \frac{\rho \text{g}}{\sigma} \left(\frac{3V}{2\pi}\right)^{2/3}, \quad \Ca = \frac{\visc\clspeed}{\sigma} 
\end{align}
are sufficiently small\footnote{Note that the length scale for the Bond number is chosen to be the radius of a volume-equivalent drop with a contact angle of \SI{90}{\degree}, i.e. $l := \sqrt[3]{3V/2\pi}$.}. However, in experiments the shape of the drop is strongly influenced by the needle which is used to deposit the drop on the substrate (see Figure~\ref{fig:deposit_water_droplet} for the deposition of a pure water droplet). While the contact line advances fast after the droplet gets in contact with the substrate, there is a complex process of detachment from the needle which is accompanied by the propagation of capillary waves and subsequent oscillations of the contact line. For partially wetting liquids ($\thetaeq > 0$) with low viscosity, the oscillations typically continue until the stationary radius is reached. Therefore, the assumption (I) is typically not applicable in this case. However, it has been reported in the literature that for highly viscous liquids such as water-glycerol mixtures, there is a viscous regime of spreading where viscous friction becomes the main source opposing capillarity \cite{Chen2014}. During this final stage of wetting after detachment from the needle, the oscillations are mainly damped out by viscosity and the drop shape is quite close to a spherical cap. Moreover, the spherical cap shape is also observed for the spreading on a polymer brush since in this case the time scale for the spreading is quite large (see Section~\ref{section:spreading_on_polymer_brush}).\\

\begin{figure}[htb]
    \centering
    \includegraphics[width=8cm]{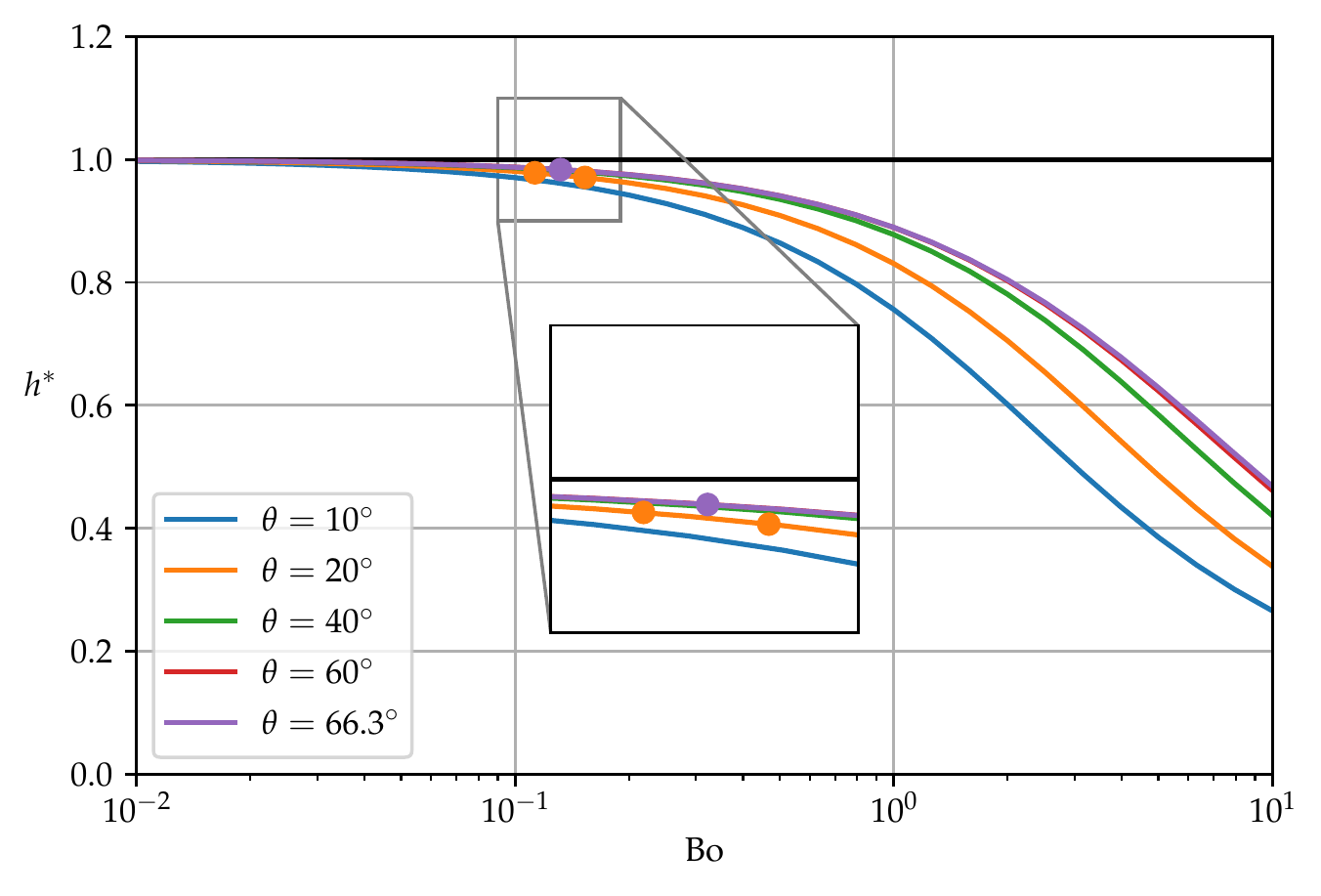}
    \caption{Comparison of dimensionless heights for varying Bond numbers and different contact angles. The curves for $\theta=\SI{60}{\degree}$ and $\theta= \SI{66.3}{\degree}$ are nearly  coinciding.}
    \label{fig:drop_shapes}
\end{figure}

In order to quantify the influence of gravity, the expected deviation of the droplet shape from a spherical cap in the \emph{stationary state} can be computed from a stationary solution of a continuum mechanical model (see Appendix~\ref{sec:appendix_mathematics} for details). We define the dimensionless height $h^\star$ of a droplet as $h^\star = h/h_{\text{cap}}(\theta,V)$, where $h_{\text{cap}}(\theta,V)$ is the height of a volume equivalent spherical cap with contact angle $\theta$. Since gravity always flattens the droplet shape, the dimensionless height in the presence of gravity satisfies
\[ 0 < h^\star < 1. \]
The algorithm described in Appendix~\ref{sec:appendix_mathematics} (see also \cite{Gruending2020}) allows to compute the full droplet shape in equilibrium and in particular the dimensionless height for varying contact angles and Bond numbers as defined in \eqref{eqn:bo_and_ca}. Figure \ref{fig:drop_shapes} shows the dimensionless drop height versus the Bond number for various contact angles. The figure displays how the dimensionless drop height decreases with increasing Bond number, e.g., due to the influence of gravity. Furthermore, the curves for the different contact angles are somewhat shifted to the right, while the distances between them decrease, meaning that the shift between curves for $\theta=\SI{20}{\degree}$ to $\theta=\SI{40}{\degree}$ is larger than the shift from $\theta=\SI{40}{\degree}$ to $\theta=\SI{60}{\degree}$. The inset shows the data points for typical drops used in the present study; see Table~\ref{tab:drop_bond_numbers} for the physical parameters. The numerical values show that the expected deviation from the height of a spherical cap is below $\SI{3}{\percent}$. 

\begin{table}[htb]
    \centering
    \begin{tabular}{l c c c c c c}
            fluid & $\theta ~ [\si{\degree}]$ & $V ~ [\si{\milli \meter \cubed}] $ & $\rho ~ [\si{\kg/\m^3}]$ & $\sigma ~ [\si{10^{-3} \frac{\newton}{\meter}}]$ &  $\text{Bo}$ & $h^\star$\\
               \hline
        water-glycerol ($75\%$) & $20.0$ & $1.0$ & $1194.9$ & $63.50$ & $0.11$ & $0.98$ \\
         water &  $20.0$ & $2.5$ & $997.05$ & $71.96$ & $0.15$ & $0.97$ \\
         water &  $66.3$ & $2.0$ & $997.05$ & $71.96$ & $0.13$ & $0.98$
    \end{tabular}
    \caption{Parameters for the comparison of the actual drop height to the spherical cap approximation.}
    \label{tab:drop_bond_numbers}
\end{table}

\paragraph{Remarks on assumption (II):} Starting from the work by Hoffman \cite{Hoffman1975} and Jiang et al.\ \cite{Jiang1979}, where it is suggested that there is a \emph{universal} relationship of the form
\begin{align}\label{eqn:contact_angle_model_v2}
\thetaapp = f(\Ca), 
\end{align}
assumption (II) is a frequent choice to model dynamic wetting, even though its validity is discussed controversially (see, e.g., \cite{Shikhmurzaev.2008}). Here $\thetaapp$ denotes the ``apparent'' (i.e.\ optically observable) contact angle and $\Ca$ is the capillary number proportional to the contact line speed $\clspeed$. Mathematically, it is more convenient to express the capillary number as a function of the contact angle, i.e.\
\begin{align}
\label{eqn:contact_angle_model}
\Ca = \psi(\thetaapp),
\end{align}
since there might be a whole interval of contact angles $[\theta_r,\theta_a]$ where the contact line does not move. This well-known effect is called contact angle hysteresis.

\begin{example}[Cox-Voinov law]
A prominent example of a relation of the type \eqref{eqn:contact_angle_model} is the Cox-Voinov law (see \cite{Bonn.2009} chapter III for a discussion) given by
\begin{align}
\mathcal{G}(\theta_{\text{app}}) = \mathcal{G}(\theta_{\text{eq}}) + \Ca \, \ln\left(\frac{x}{L}\right).
\end{align}
Here $x/L$ is the ratio of macroscopic to microscopic length scales, an unknown parameter that has to be found by fitting experimental data (Bonn et al.\ \cite{Bonn.2009} report $x/L=10^4$. The function $\mathcal{G}$ defined as
\[ \mathcal{G}(\theta) = \int_0^\theta \frac{x - \sin x \cos x}{2 \sin x} \, dx \]
can be well approximated by $\theta^3/9$ for $\theta < 135^\circ$ (see \cite{Bonn.2009}), so that in the following we will refer to the simplified equation
\begin{align}
\label{eqn:cox-voinov}
\theta_{\text{app}}^3 = \theta_{\text{eq}}^3 + 9 \Ca \, \ln\left(\frac{x}{L}\right)
\end{align}
as the ``Cox-Voinov law''. Equation~\eqref{eqn:cox-voinov} results from an asymptotic solution of a hydrodynamic model by Cox~\cite{Cox.1986}, which is valid in the limit $\Ca, \text{Re} \rightarrow 0$ for an ideal surface with no heterogeneities, if the microscopic contact angle equals the equilibrium contact angle $\thetaeq$. According to \cite{Blake.2006}, the approximation $\theta_m \approx \thetaeq$ is justified if viscous dissipation is large compared to local dissipation at the contact line.
\end{example}

\begin{example}[Kistler's empirical function]
Another commonly used example is the empirical function by Kistler \cite{kistler1993} given by the expression
\begin{equation}
\label{eqn:hoffman_kistler_model}
\begin{aligned}
\thetaapp &= f_{\text{Hoff}}(\Ca) \\
&= \cos^{-1}\left(1 - 2 \tanh\left[ 5.16 \left(\frac{\Ca}{1+ 1.31 \Ca^{0.99}} \right)^{0.706} \right]\right).
\end{aligned}
\end{equation}
The latter function has been obtained from a fit to the experimental data by Hoffman \cite{Hoffman1975}. In order to describe partial wetting, the function is shifted according to
\begin{align} 
\label{eqn:hoffman_kistler_model_partial_wetting}
\thetaapp = f_{\text{Hoff}} [ \Ca + f_{\text{Hoff}}^{-1}(\theta_0) ]. 
\end{align}
\end{example}

\begin{figure}[hbt]
 \centering
 \includegraphics[width=8cm]{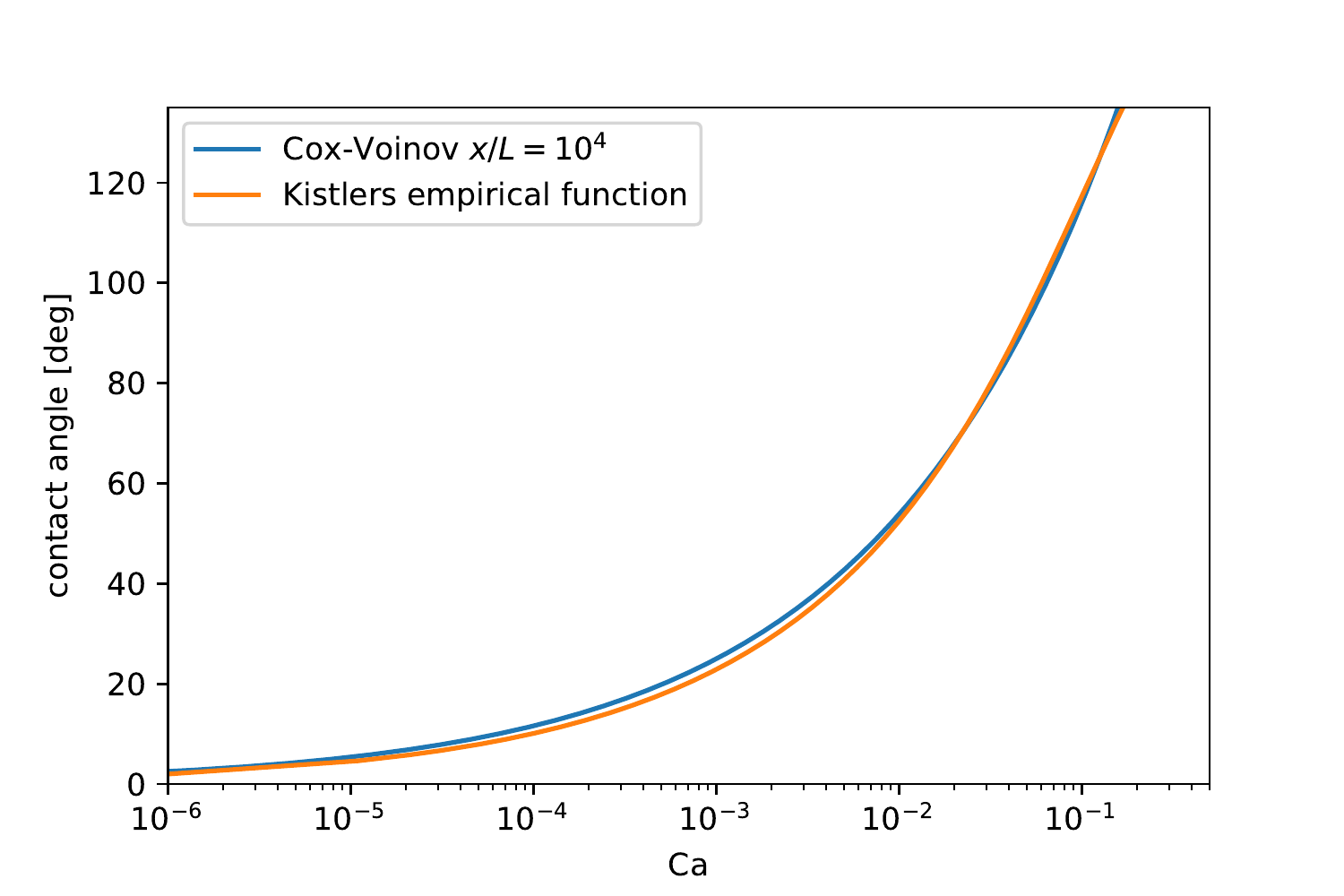}
 \caption{Kistler's empirical function and the Cox-Voinov law for $\thetaeq = 0$.}
 \label{fig:hoffman_cox_voinov}
\end{figure}

A direct comparison of the empirical models \eqref{eqn:cox-voinov} for $x/L = 10^4$ and \eqref{eqn:hoffman_kistler_model_partial_wetting} in the case of complete wetting shows that the two models are very similar; see Figure~\ref{fig:hoffman_cox_voinov}. Both empirical relations have been shown to be compatible with experimental data over a range of capillary numbers for different fluid/substrate combinations.

\section{Spreading Dynamics of Spherical Drops}
\label{section:theory}
\subsection{Spreading dynamics in phase space}
It can be advantageous to consider the spreading dynamics in phase space (see \cite{Hartmann.2020} for the phase space diagram of a breakup process). The idea is to express the time-derivative of the base radius $\dot{L}$ as a function of the base radius itself. Formally, we define
\begin{align}
\dot{L}(L_0) := \ddt{L}\left(L^{-1}(L_0)\right).
\end{align}
Note that this is possible since $L(t)$ is a monotonically increasing function with inverse $L^{-1}$. The advantage of the latter approach is that it eliminates the dependence of the data on the arbitrary choice of $t_0$. This is helpful since the instant of contact $t_0$ with $L(t_0) = 0$ might, in practice, be hard to determine experimentally with precision. In fact, the function $\dot{L}(L)$ is invariant with respect to a shift in the time coordinate. It is easy to show that if $L$ is given by
\[ L(t) = C (t-t_0)^\beta \]
for some $C, \, \beta > 0$ and $t_0 \in \RR$, then
\[ \dot{L}(L) = (\beta C^{1/\beta}) \, L^{1-1/\beta}. \]
For the famous Tanner law \cite{Tanner_1979} ($\beta = 0.1$) this means $\dot{L}(L) \propto L^{-9}$.

\subsection{Geometry-based model for spreading drops}
\begin{figure}[ht]
\subfigure[Notation.]{\includegraphics[width=6cm]{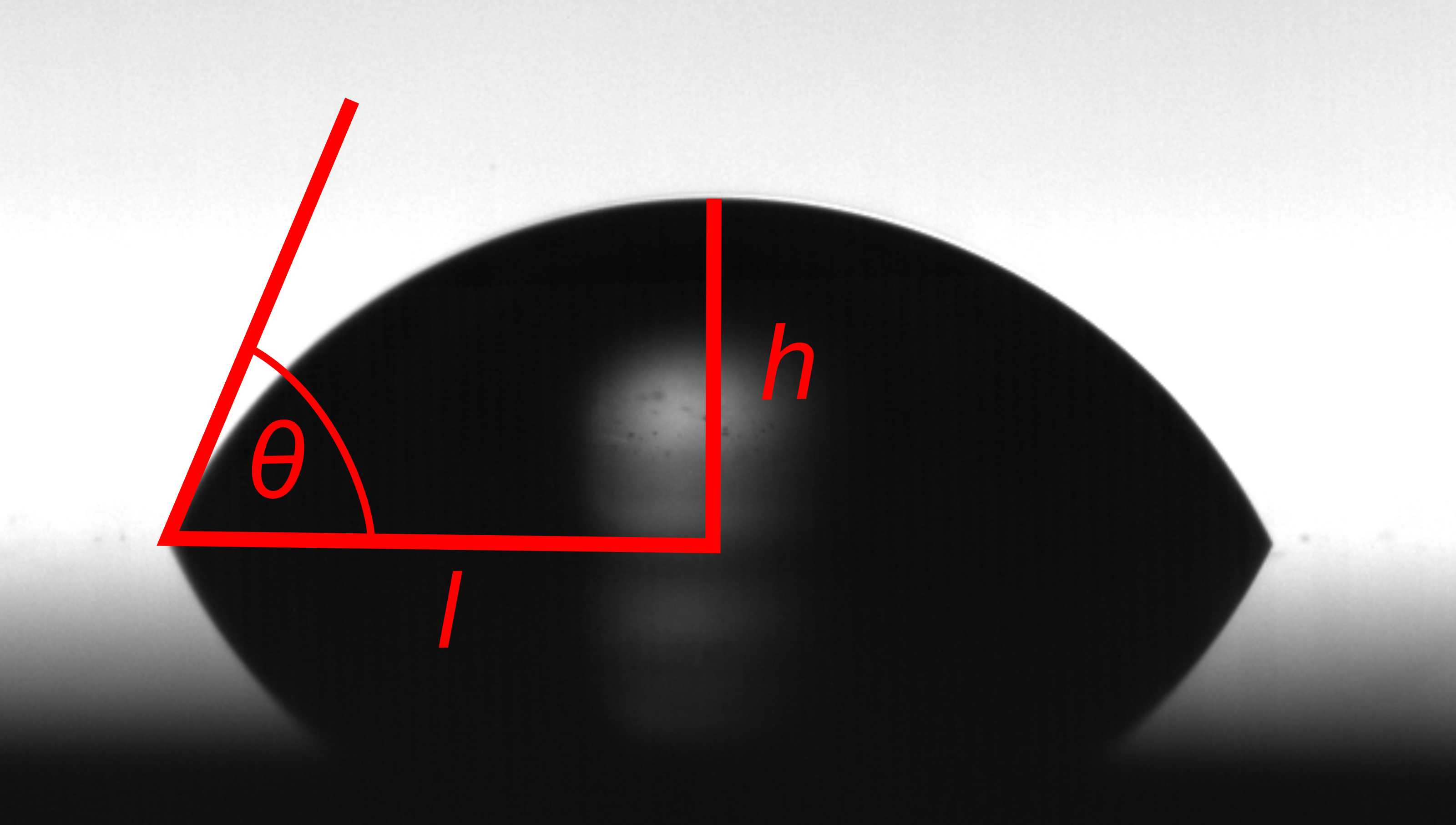}\label{fig:notation}}
\subfigure[The geometric relation $g$.]{\includegraphics[width=6cm]{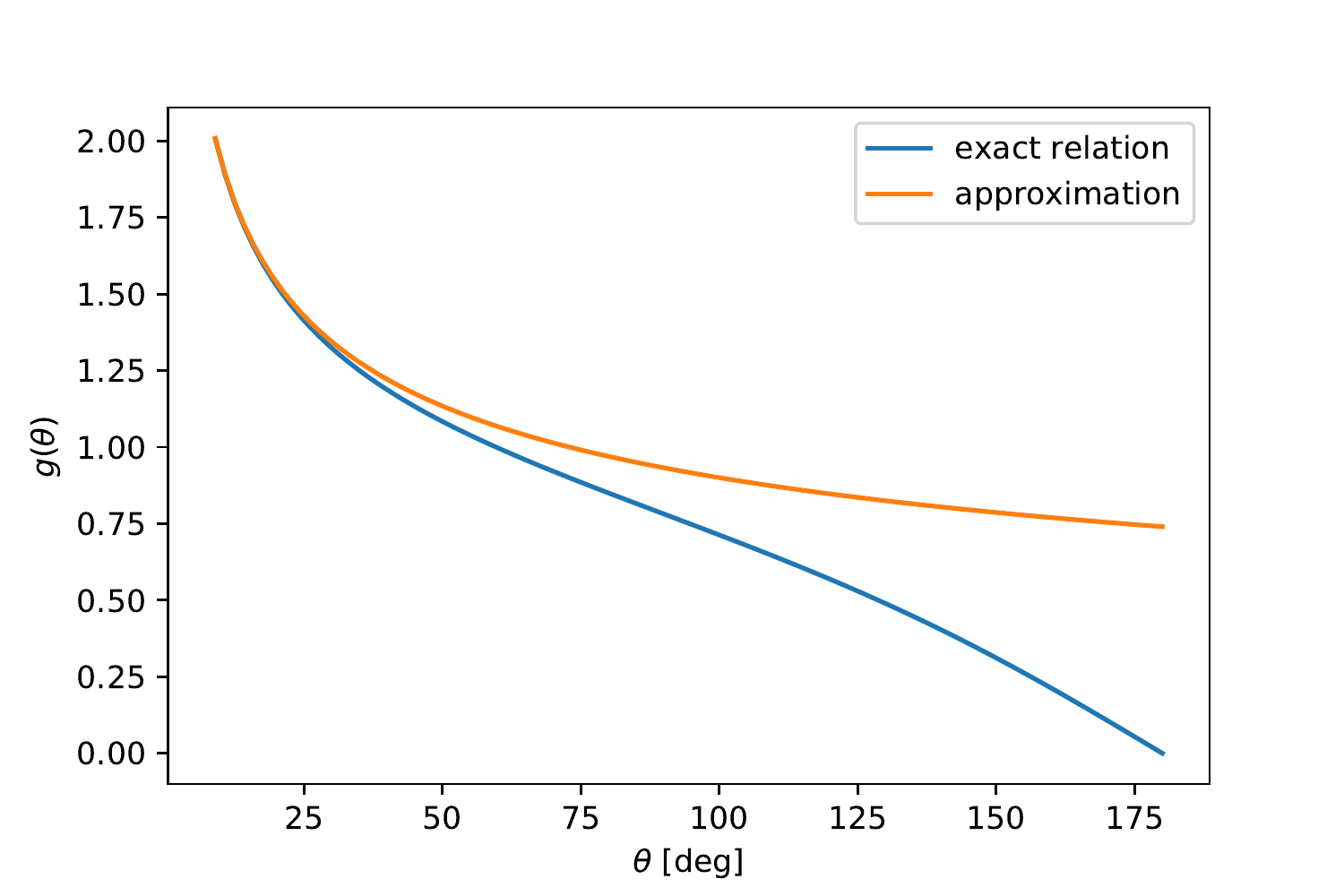}\label{fig:geometric_relation}}
\caption{Some basic geometry.}
\end{figure}

For a \emph{spherical} droplet, there is a purely geometric relationship between the base-radius $L$, the volume $V$ and the contact angle according to
\begin{align}
\label{eqn:basic_geometric_relation}
\frac{L}{V^{1/3}} = g(\theta) := \sin\theta \left(\frac{\pi(1-\cos\theta)^2(2+\cos\theta)}{3}\right)^{-1/3};
\end{align}
see Fig.~\ref{fig:notation} for notation.
Note that $g$ is monotonically decreasing with $\theta$ (see Figure~\ref{fig:geometric_relation}) and hence invertible on $[0,\infty)$. For small values of $\theta$, i.e. for flat drops, the function $g$ can be approximated by (see~\cite{Gennes.1985})
\[ g(\theta) \approx \hat{g}(\theta) = \left( \frac{4}{\pi \theta} \right)^{1/3}. \]
Inverting relation \eqref{eqn:basic_geometric_relation} allows to express the contact angle as a function of the base radius and the volume according to
\begin{align} 
\label{eqn:geometric_theta}
\theta = g^{-1}\left( \frac{L}{V^{1/3}} \right). 
\end{align}
Using the approximation for flat drops allows to simplify the above equation as
\begin{align}
\label{eqn:geometric_theta_approximation}
\theta = \frac{4}{\pi} \frac{V}{L^3}. 
\end{align}
Assuming that the contact angle is related to the contact line speed $\clspeed$ by an empirical relation of the type \eqref{eqn:contact_angle_model} allows to derive an ordinary differential equation for the spreading dynamics of spherical drops. Note that this observation has already been made by de Gennes~\cite{Gennes.1985}, where (in the case of complete wetting) the approximation \eqref{eqn:geometric_theta_approximation} for flat drops is applied. Here we generalize this idea to arbitrary contact angles.\\
\\
Since $\dot{L}=\clspeed$, it follows from \eqref{eqn:geometric_theta} and \eqref{eqn:contact_angle_model} that the base radius satisfies the ordinary differential equation
\begin{align}
\label{eqn:base_radius_evolution}
\frac{\eta}{\sigma} \, \dot{L}(t) = \psi\left(g^{-1}\left( \frac{L(t)}{V(t)^{1/3}}\right)\right), \quad L(t_0) = L_0.
\end{align}

\begin{remark} From the form of equation~\eqref{eqn:base_radius_evolution} we draw the following conclusions:
\begin{enumerate}[(i)]
 \item The problem is uniquely solvable provided that $\psi$ is, e.g., Lipschitz continuous. 
 \item Since thermodynamics of moving contact lines implies (see,e.g.,\cite{Ren.2010,Fricke.2019})
\[ \psi \geq 0 \quad \text{for} \quad \theta \geq \thetaeq, \]
it follows directly that (since $g^{-1}$ is monotonically decreasing)
\[ \dot{L} \geq 0 \quad \text{if} \quad L \leq V^{1/3} g(\thetaeq) =: \eqlength(\thetaeq,V)  \]
 and vice versa for $L\geq\eqlength$.
 \item The ordinary differential equation \eqref{eqn:base_radius_evolution} is autonomous if the volume is conserved. However, this is not a necessary assumption. The dynamic volume can be incorporated if the function $V=V(t)$ is known.
 \item The dynamics of the contact angle can be inferred from a solution of \eqref{eqn:base_radius_evolution}, using the relation \eqref{eqn:geometric_theta}. Moreover, one can also derive an equivalent  evolution equation for the contact angle, see below.
\end{enumerate}
\end{remark}

\begin{mdframed}
By plotting the experimental data for  $\Ca = \frac{\visc}{\sigma} \dot{L}$ vs. $\theta(L,V)=g^{-1}(L/V^{1/3})$ one can directly read off an \emph{empirical function} $\psi$, provided that the droplet is spherical and the volume $V$ is known.
\end{mdframed}

\textbf{Note:} The above method delivers an empirical function $\psi$ describing the data of an individual experiment even in the case when assumption (II) does \emph{not} hold. For example, the function $\psi$ could depend on further variables like temperature
\begin{align}
\Ca = \psi(\theta,T,\dots).
\end{align}
In fact, experiments in curtain coating \cite{Blake1999} suggest that $\psi$ may also depend on the flow field near the contact line - an effect known as ``hydrodynamic assist''. Therefore, the proposed method should be understood as a tool to probe whether or not the data can be collapsed onto a single curve for $\psi$ (at least for some range of parameters). It is only in the latter case, that  Equation~\eqref{eqn:base_radius_evolution} is able to \emph{predict} the spreading dynamics of spherical droplets (see Section~\ref{sec:results}).

\paragraph{Thin droplet approximation:} Using the approximation \eqref{eqn:geometric_theta_approximation} for small contact angles, one can approximate \eqref{eqn:base_radius_evolution} by the ODE
\begin{align}
\label{eqn:base_radius_evolution_small_theta}
\dot{L}(t) = \frac{\sigma}{\visc} \, \psi\left(\frac{4}{\pi} \frac{V(t)}{L(t)^3} \right).
\end{align}
If the volume is constant and $\psi$ is given by a power law (corresponding to complete wetting), i.e.\
\[ \Ca = c \, \theta^m,  \]
then \eqref{eqn:base_radius_evolution_small_theta} reads (see \cite{Gennes.1985})
\[ \dot{L} = \tilde{c} \, V_0^m L^{-3m}. \]
In case $L(0)=0$, the evolution of the base radius obeys the power law
\[ L \propto V_0^{\frac{m}{3m+1}} \, t^{\frac{1}{3m+1}}. \]
Hence the Tanner law $L \propto t^{1/10}$ is obtained for $m=3$.

\paragraph{Contact angle dynamics:} Differentiating the geometrical relation \eqref{eqn:basic_geometric_relation} with respect to time yields
\begin{align}
\label{eqn:dynamic_geometric_relation}
\dot{L} = V^{1/3} g'(\theta) \, \dot\theta + g(\theta) \, \ddt{} V^{1/3}.
\end{align}
where $\clspeed = \dot{L}$ denotes the contact line speed. Then equation \eqref{eqn:dynamic_geometric_relation} combined with \eqref{eqn:contact_angle_model} yields the ordinary differential equation
\begin{align}
\label{eqn:theta_evolution}
\dot\theta = \frac{\frac{\sigma}{\visc} \, \psi(\theta)-g(\theta)\ddt{} V^{1/3}}{g'(\theta) V^{1/3} }, \quad \theta(t_0) = \theta_0.
\end{align}

\paragraph{Non-dimensional form for a constant volume:} For constant volume, i.e.\ $V(t) \equiv V_0$, one may choose $\eqlength = V_0^{1/3} g(\thetaeq)$ as a length scale and 
\[ \tau = \frac{\visc V_0^{1/3}}{\sigma} \]
as a time scale for non-dimensionalization according to
\[ \tilde{L}(\tilde{t}) = \frac{L(\tau \tilde{t})}{\eqlength} \quad \Leftrightarrow \quad L(t) = \eqlength \tilde{L}\left(\frac{t}{\tau} \right). \]
Then equations \eqref{eqn:base_radius_evolution} becomes
\begin{align}
\tilde{L}' = \frac{1}{g(\thetaeq)} \psi\left(g^{-1}\left[g(\thetaeq) \, \tilde{L} \right]\right).
\end{align}
Moreover, the non-dimensional form of $\theta(t)$, defined as
\[ \tilde{\theta}(\tilde{t}) = \theta(\tau \tilde{t}),  \]
satisfies the ordinary differential equation
\begin{align}
\tilde{\theta}' = \frac{\psi(\tilde{\theta})}{g'(\tilde{\theta})}. 
\end{align}

\section{Experimental Methods}
\begin{figure}[hbt]
 \centering
 \includegraphics[width=6.25cm]{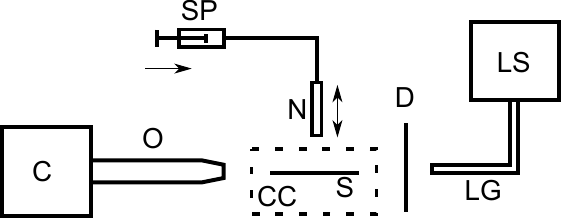}
 \caption{Experimental setup: A drop is dosed with a syringe pump (SP) which is connected to a needle (N) and placed on the substrate (S) which is eventually placed inside a climate chamber (CC). The spreading process is recorded with a high-speed camera (C) connected to a ling distance microscope (O). Illumination is performed with a cold light source (LS) from which light is lead with a light guide (LG) to a diffuser (D).}
 \label{fig:experimentalSetup}
\end{figure}

In order to validate the theoretical predictions, spreading experiments are performed:
To enable a precise dosing with defined flow rates, the setup is equipped with a gas tight glass syringe (Hamilton, SYR \SI{10}{\milli\liter}, 1010TLL, no STOP) placed in a syringe pump (Legato 100, KD Scientific) and attached to a FEP tube with a metal needle (Hamilton, Teflon tip, GA 26) at its end.
In order to assure slow movement of the needle perpendicular to the substrate, the needle is fixed in a self-built construction set onto a z-stage which can be moved manually and independently from the rest of the setup.
Imaging is performed with a high-speed camera (Photron SA-X1) at 10,000 fps.
A Navitar 12-X long distance microscope attached to the camera assures a sufficient spatial resolution.
Back-light illumination is performed with a cold light source (Volpi intraLED 5).
Diffusive light is generated by placing a self-built diffuser in front of the light guide which comes from the light source.
Camera and objective, the needle tip and the diffuser are arranged in a way that they lie on one axis (Fig.~\ref{fig:experimentalSetup}).\\

Spreading experiments are performed by first producing a pendant drop at the needle tip. The height of the needle is then decreased slowly, so that the droplet comes into contact with the substrate. The spreading process of the sessile drop is then recorded with the high-speed camera. In the present study, we consider two different spreading experiments, namely
\begin{enumerate}[(i)]
 \item the spreading of a 75~wt.\% glycerol-water droplet on a bare silicon wafer and
 \item the spreading of a pure water droplet on a PNIPAm coated Si wafer.
\end{enumerate}
In the first case, the volume of the droplets is approximately 1~$\mu$L. The droplets consist of 75~wt.\% glycerol (Sigma-Aldrich,  $\geq$99.5\%,CAS: 56-81-5) in water (Milli-Q, 18.2~M$\Omega$ cm s). Since the spreading process is much faster than the evaporation process, the atmosphere around the droplet is not controlled.\\
\\
This, on the other hand, is the case when the spreading kinetics of 2~$\mu$L water droplets on PNIPAm coated Si substrates are measured. The relative humidity is varied between 15-80~\% and the long time spreading process is recorded at 50~fps for 3~min. For further information on the preparation of the PNIPAm coated substrates, the reader is referred to Appendix~\ref{sec:polymer_brush_preparation}. Humidification is performed by leading a volume stream of nitrogen through a bubbler which contains DI-water and mixing it with an unhumidified nitrogen stream. By independently setting the flow rates of the humidified and unhumidified nitrogen, different relative humidities can be achieved inside a climate chamber, in which the mixed stream is leaded. The needle of the self-built dosing system can penetrate the climate chamber by stitching through parafilm, which serves as a sealing above the substrate inside the chamber.\\
\\
All experiments are evaluated with the inhouse \textit{Matlab}-algorithm called \textit{Drop of EvolutioN: Impact, Imbibition, Propagation, evaporation (DENIISE)}. The algorithm processes the two-dimensional images from the high-speed camera and delivers (assuming an axisymmetric shape) the foot radius $L$, the height $h$, the volume $V$ and the contact angle $\theta$ for each frame.
 
\section{Experimental Results}
\label{sec:results}
\subsection{Spreading of viscous droplets on a homogeneous solid substrate}
\label{sec:glycerol_75}
\begin{figure}[htb]
\subfigure[Experimental data for the drop volume.]{\includegraphics[width=6.5cm]{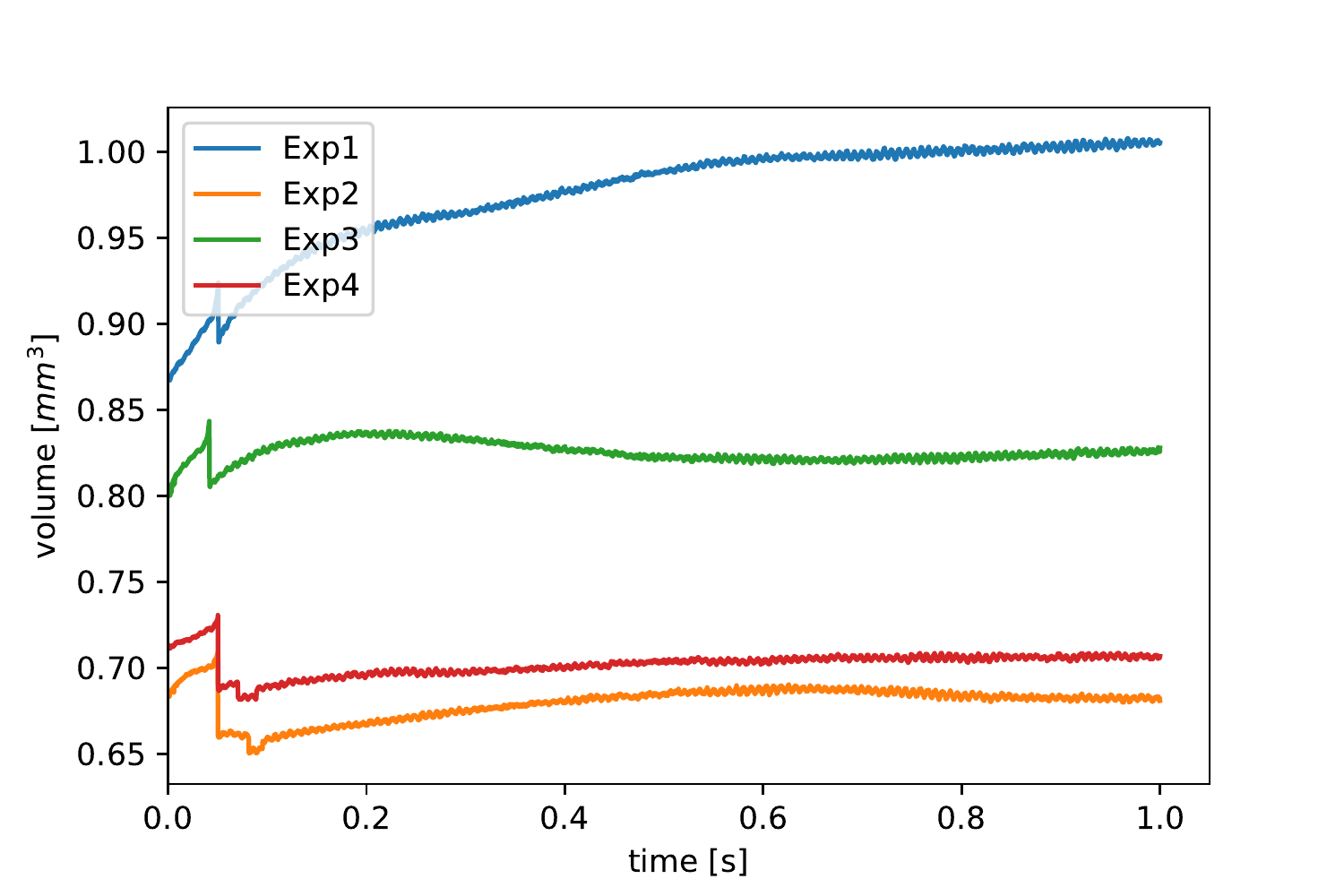}\label{fig:glycerol_75/volume}}
 \subfigure[Experimental data for the foot radius.]{\includegraphics[width=6.5cm]{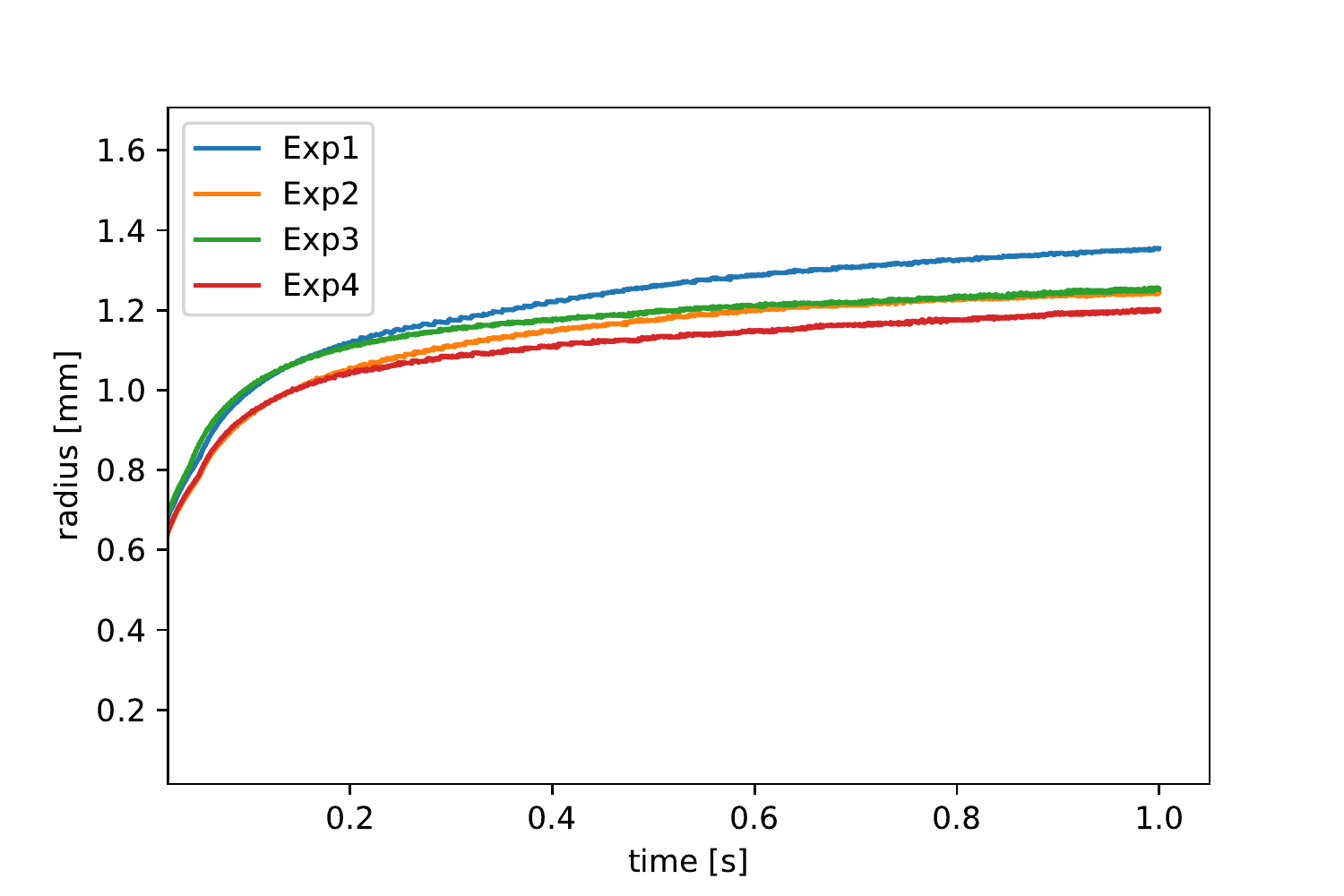}\label{fig:glycerol_75/radius}}
 \caption{Spreading of a $75\%$-glycerol droplet on a silicon wafer.}
\end{figure}

We consider the spreading of water-glycerol droplet ($75\%$ glycerol) on a bare silicon wafer. The high dynamic viscosity $\visc = 29.96 \, \text{m} \text{Pa} \cdot \text{s}$ of the liquid leads to viscous stage of spreading where the droplet spreads as a spherical cap (see \cite{Chen2014}). Note that the viscosity is increased with respect to pure water by a factor of more than $30$. Experimental data for the spreading dynamics have been recorded for $4$ repetitions of the experiment. Fig.~\ref{fig:glycerol_75/volume} shows the experimental data for the drop volume as a function of time. Note that the process of release from the needle is visible as a kink in the data. Apparently, this type of droplet application leads to a variation in the drop volume of approximately $35\%$. The latter also leads to different equilibrium foot lengths of the individual drops; see Fig.~\ref{fig:glycerol_75/radius}.\\

\begin{figure}[htb]
\centering
\includegraphics[width=8.0cm]{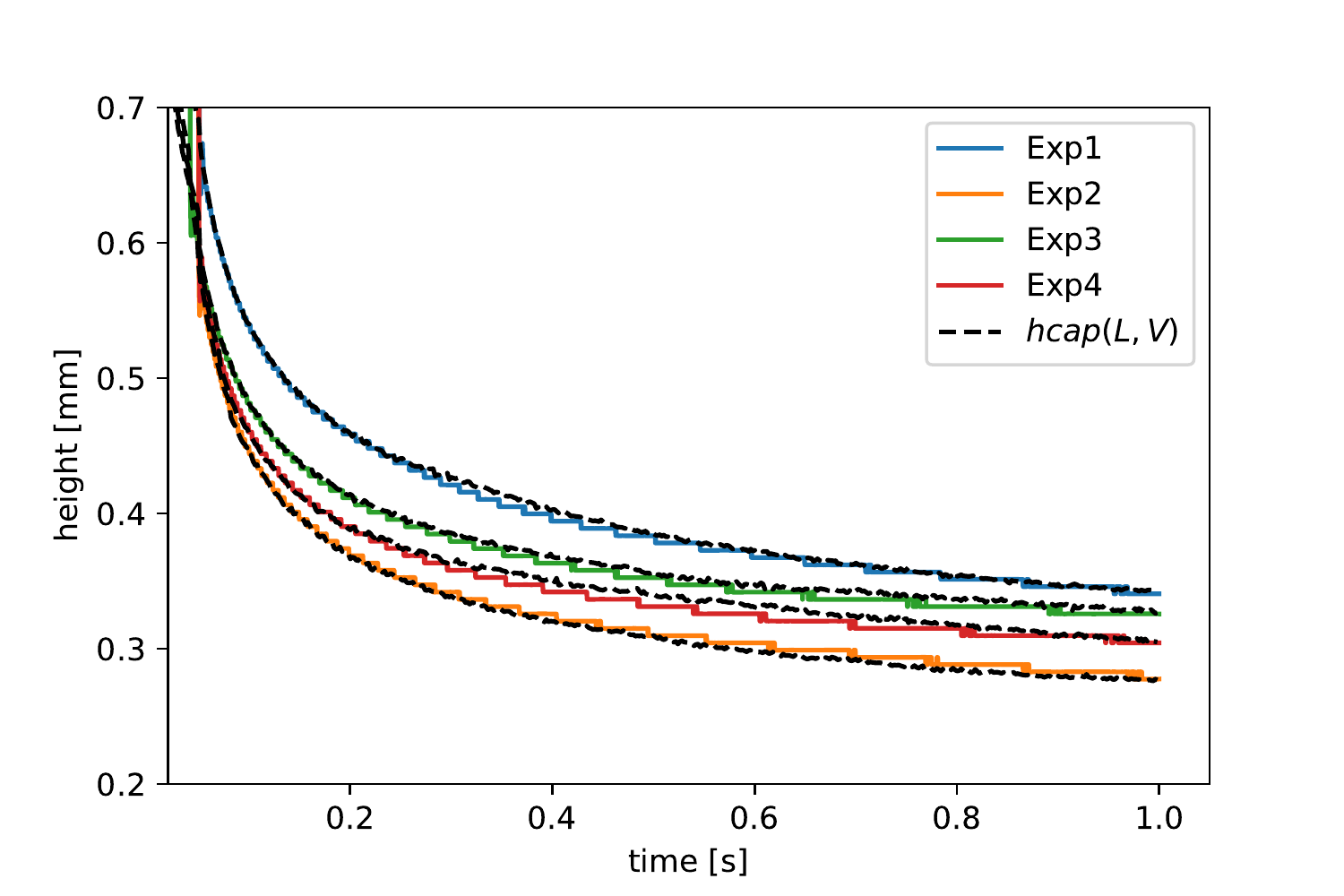}
\caption{Experimental data for the height compared to $h_{\text{cap}}(L_{\text{exp}},V_{\text{exp}})$.}
\label{fig:glycerol_75/height}
\end{figure}

\begin{figure}[htb]
\subfigure[Experiments 1 and 2.]{\includegraphics[width=6.5cm]{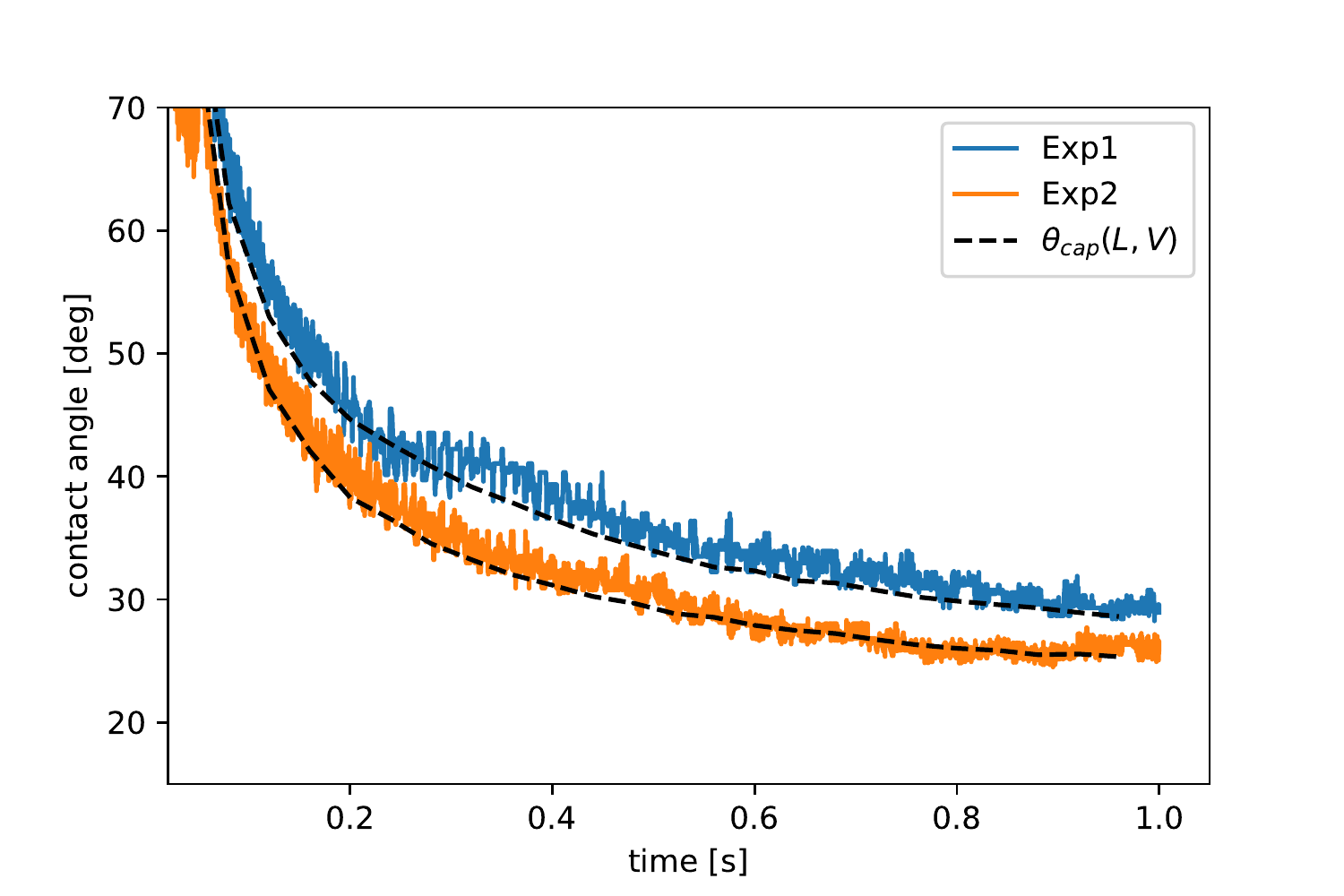}\label{fig:glycerol_75/contact_angle_exp1_exp2}}
\subfigure[Experiments 3 and 4.]{\includegraphics[width=6.5cm]{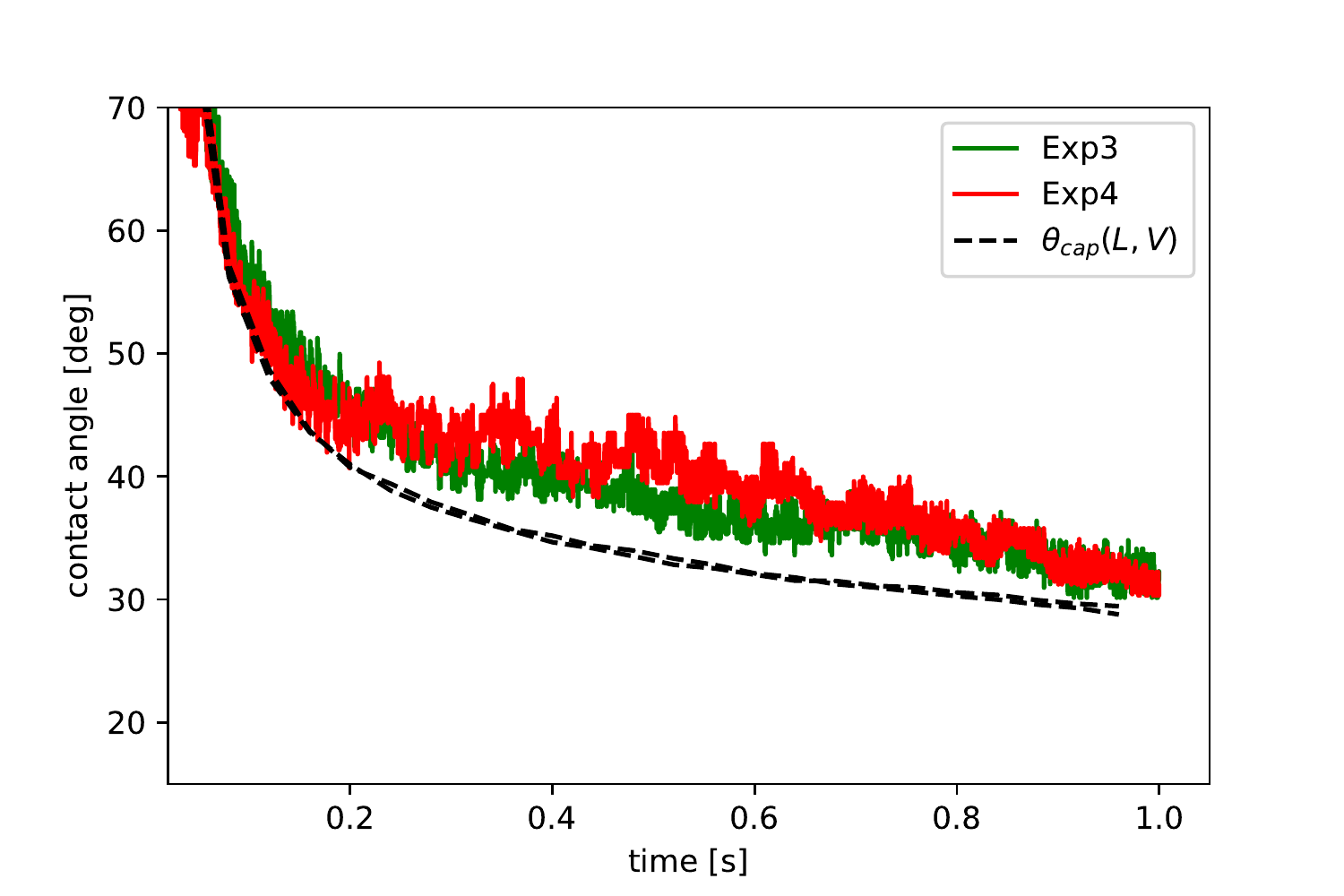}\label{fig:glycerol_75/contact_angle_exp3_exp4}}
\caption{Experimental data for the contact angle.}
\end{figure}

The consistency of the experimental data with respect to the spherical cap approximation can be checked in two ways, namely
\begin{enumerate}[(i)]
 \item by comparing the experimental drop height $h_{\text{exp}}$ with $h_{\text{cap}}(L_{\text{exp}},V_{\text{exp}})$ and
 \item by comparing the experimental contact angle with $\theta_{\text{cap}}(L_{\text{exp}},V_{\text{exp}})$.
\end{enumerate}
The experimental data for the drop height for all experiments agree well with the spherical cap approximation; see Fig.~\ref{fig:glycerol_75/height}. According to the results in Fig.~\ref{fig:glycerol_75/contact_angle_exp1_exp2}, there is also a reasonable agreement of the contact angle with the theoretical values for a spherical cap for Experiments 1 and 2. However, we observe a systematic deviation in the contact angle for the Experiments 3 and 4; see Fig.~\ref{fig:glycerol_75/contact_angle_exp3_exp4}.\\

\begin{figure}[htb]
\subfigure[]{\includegraphics[width=6.5cm]{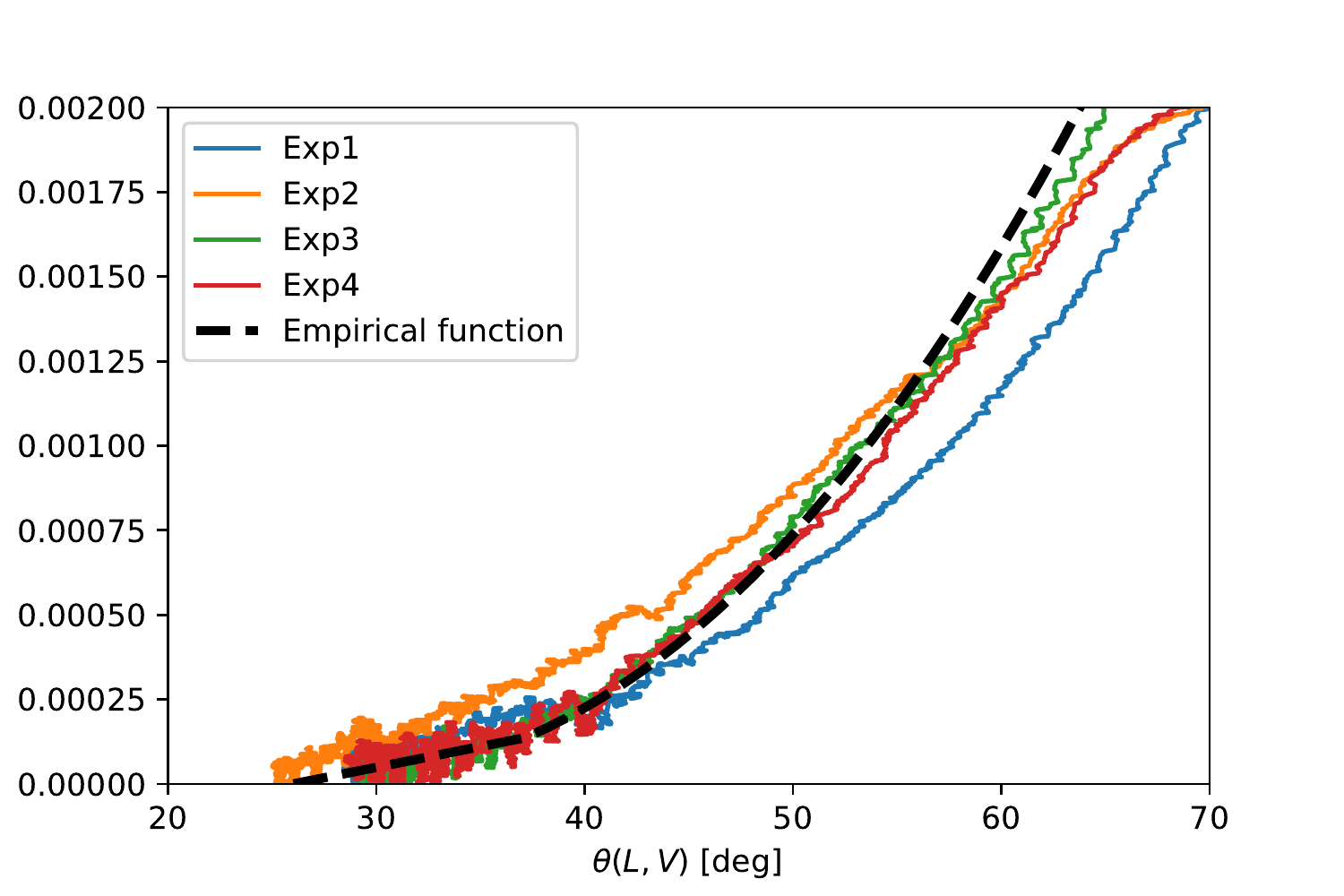}\label{fig:glycerol_75/individual_psi}}
\subfigure[]{\includegraphics[width=6.5cm]{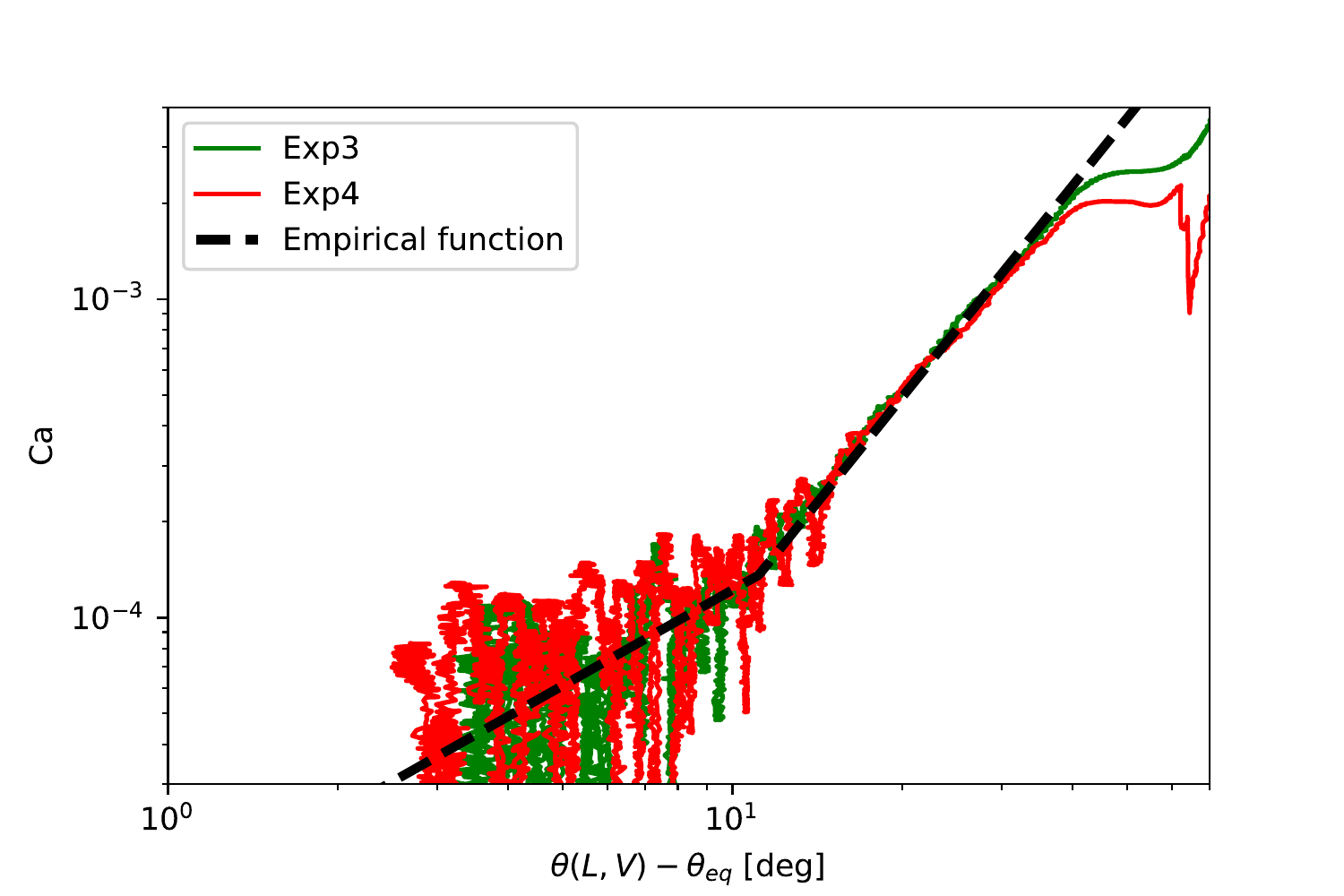} \label{fig:glycerol_75/psi_loglog}}
\caption{Empirical function $\psi$.}
\end{figure}

\begin{figure}[htb]
\subfigure[ODE solution for the base radius.]{\includegraphics[width=6.5cm]{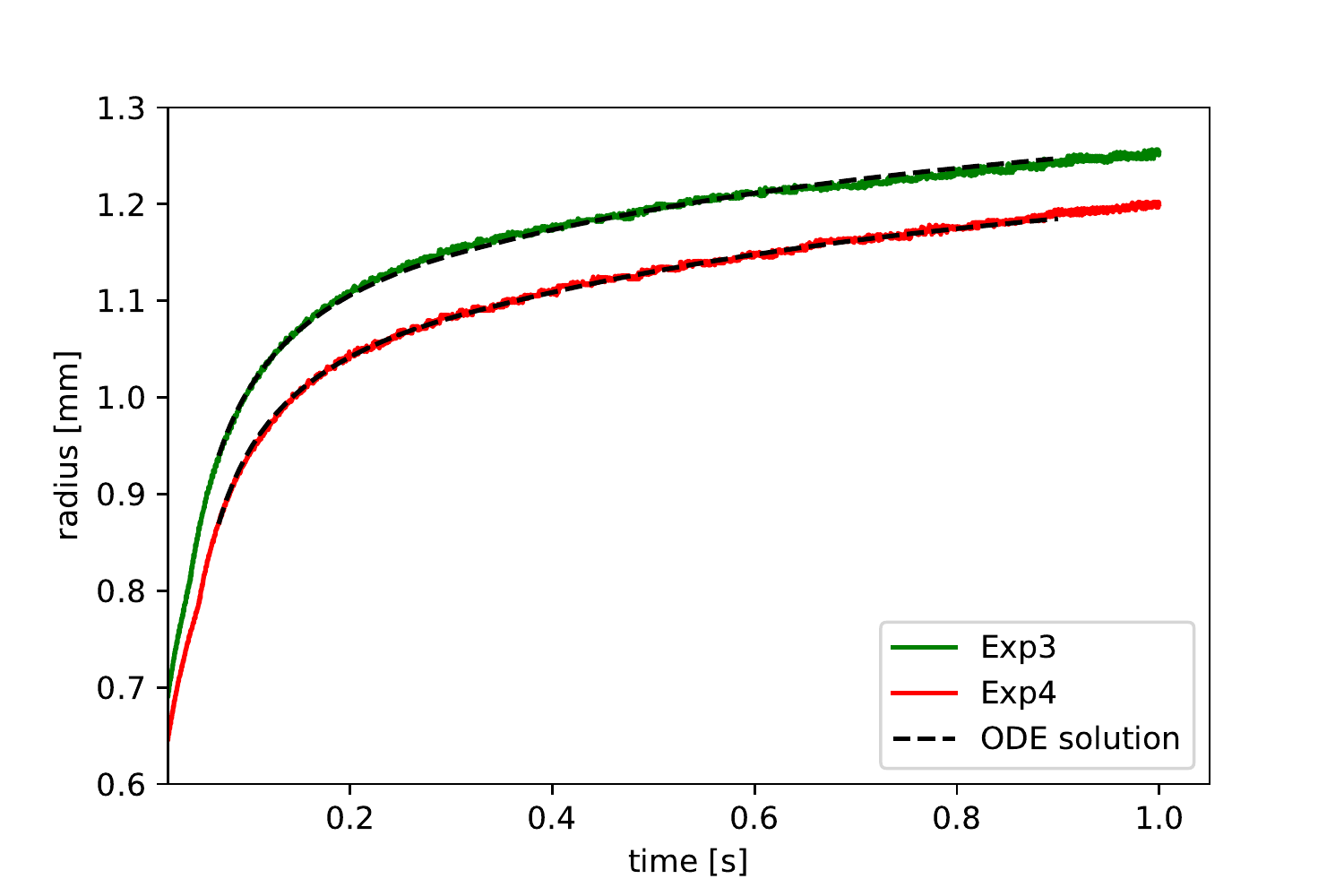}\label{fig:glycerol_75/solution_radius}}
\subfigure[Comparison with literature.]{\includegraphics[width=6.5cm]{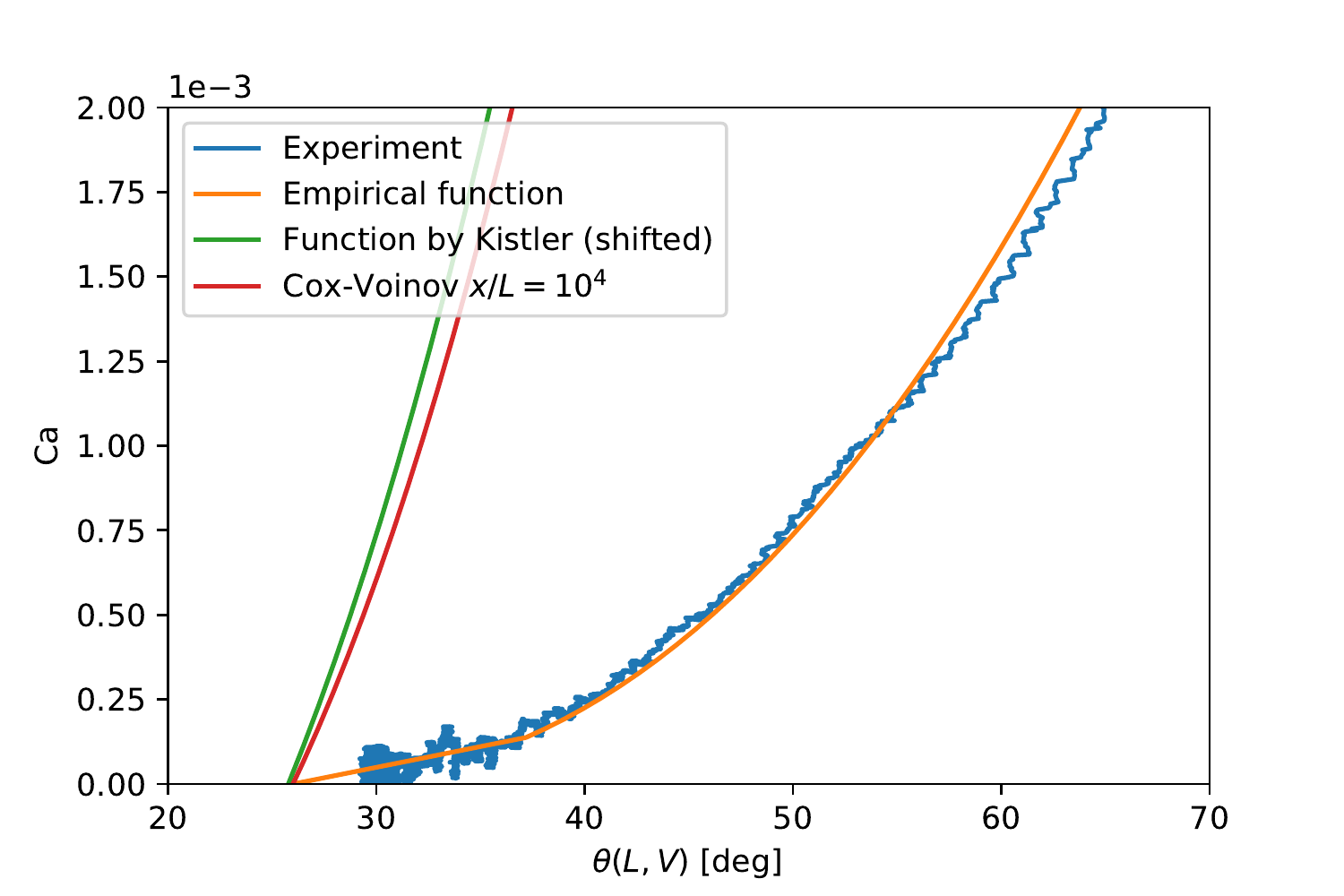} \label{fig:glycerol_75/literature_comparison}}
\caption{ODE solution and literature comparison.}
\end{figure}

The contact line velocity is obtained from the numerical differentiation of the experimental data for the base radius with respect to time. The resulting capillary number is plotted against the contact angle \emph{computed} as a function of the measured base radius $L$ and volume, i.e.\ according to eq.~\eqref{eqn:geometric_theta}; see Fig.~\ref{fig:glycerol_75/individual_psi}. Note that this method leads to a much smaller scatter in the data since the base radius and the volume can be measured with higher precision than the contact angle itself. We find that the experimental data in the $\theta$-$\Ca$ plane are quite close for Experiment 3 and 4 while Experiments 1 and 2 show some offset. The data for Experiment 3 and 4 can be described by the common empirical function
\begin{align}
\label{eqn:empirical_psi_homogeneous_surface}
\psi_{\text{emp}}(\theta) = \max\{5 \cdot 10^{-3} (\theta-\thetaeq)^{2.2}, \, 7 \cdot 10^{-4} (\theta-\thetaeq) \},
\end{align}
where $\thetaeq = 26.0^\circ$. Even though noise in the data leads to an oscillatory signal for low capillary numbers, the data show that the empirical function cannot be described by one single exponent; see Fig.\ref{fig:glycerol_75/psi_loglog}. The solution of the ODE \eqref{eqn:base_radius_evolution} with $\psi$ given by \eqref{eqn:empirical_psi_homogeneous_surface} is plotted in Fig.~\ref{fig:glycerol_75/solution_radius}. Here, $V=V(t)$ is computed from the experimental data for the volume by linear interpolation, and the initial conditions for the ODE are taken from the experimental data after detachment from the needle. It is observed that the dynamics for both Experiment 3 and 4 agree very well with the ODE solution employing the common empirical function \eqref{eqn:empirical_psi_homogeneous_surface}.

\paragraph{Comparison with literature relations:} It is remarkable to note that there is a large discrepancy between the empirical function \eqref{eqn:empirical_psi_homogeneous_surface} which describes the present experiment and the well-known relations shown in Figure~\ref{fig:glycerol_75/literature_comparison}. Both the Cox-Voinov relation \eqref{eqn:cox-voinov} with $x/L=10^4$ as reported in \cite{Bonn.2009} and the Kistler function given by \eqref{eqn:hoffman_kistler_model_partial_wetting} are far off from the present data. Apparently, the Cox-Voinov law appears to not apply for the present case, even though viscous dissipation is large and the capillary number is small. This discrepancy might be related to the polarity of the fluid or to mixture effects; see also \cite{Hayes1993} where it is stated that the hydrodynamic model \eqref{eqn:cox-voinov} is unable to produce physically reasonable values for the slip length $L$ for polar liquids on a PET surface.

\subsection{Spreading of water droplets on a swellable polymer brush}
\label{section:spreading_on_polymer_brush}
We consider water drops with volume $V \approx 2 \, \text{mm}^3$ under standard conditions  ($T=25^\circ \text{C},\ p = 1 \ \text{bar}$) slowly spreading ($\clspeed \approx 10 \mum/s$) on a PNIPAm polymer brush (see Appendix~\ref{sec:polymer_brush_preparation} for details about the polymer brush). Typical values for the Bond and Capillary number are
\[ \text{Bo} = \frac{\rho \text{g}}{\sigma} \left(\frac{3V}{2\pi}\right)^{2/3} \approx 0.13, \quad \Ca = \frac{\visc\clspeed}{\sigma} \approx 1.2 \cdot 10^{-7}. \]
The shape of the droplet after detachment from the needle is close to a spherical cap. Moreover, the droplet retains a spherical shape throughout the whole process while the volume change is mainly due to evaporation. Note, a part of the drop volume is transported beyond the contact line into the thin polymer film. However, given the small heights of the swollen brush (less than 300 nm), this volume may be neglected in regard to an overall change in volume of the drop. Like in the previous experiments, the volume of the drop is extracted from the images of the high-speed camera.

\begin{figure}[htb]
\subfigure[Experimental data for the drop volume.]{\includegraphics[width=6.5cm]{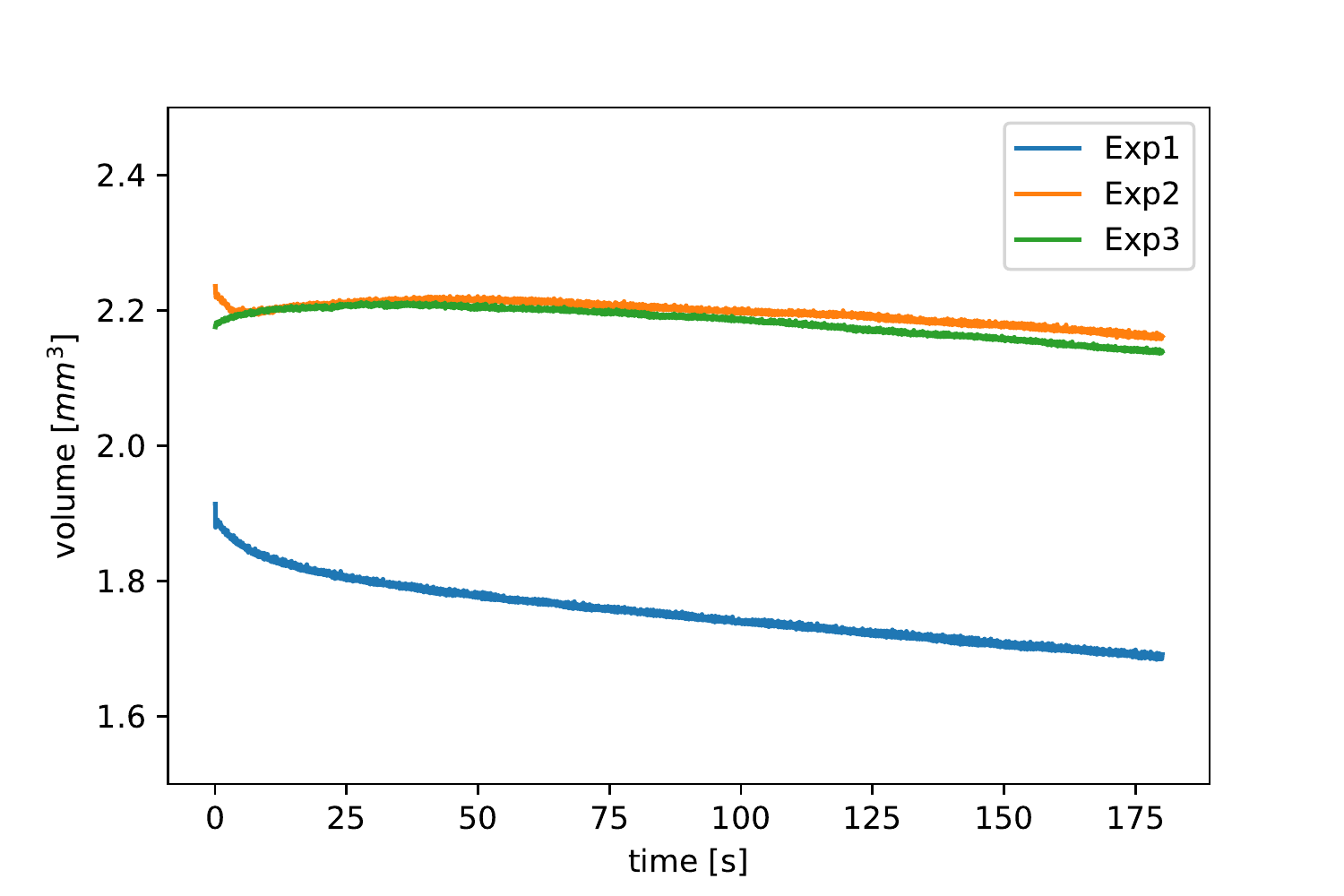}\label{fig:polymer-brush_80/volume}}
\subfigure[Experimental data for the drop base radius.]{\includegraphics[width=6.5cm]{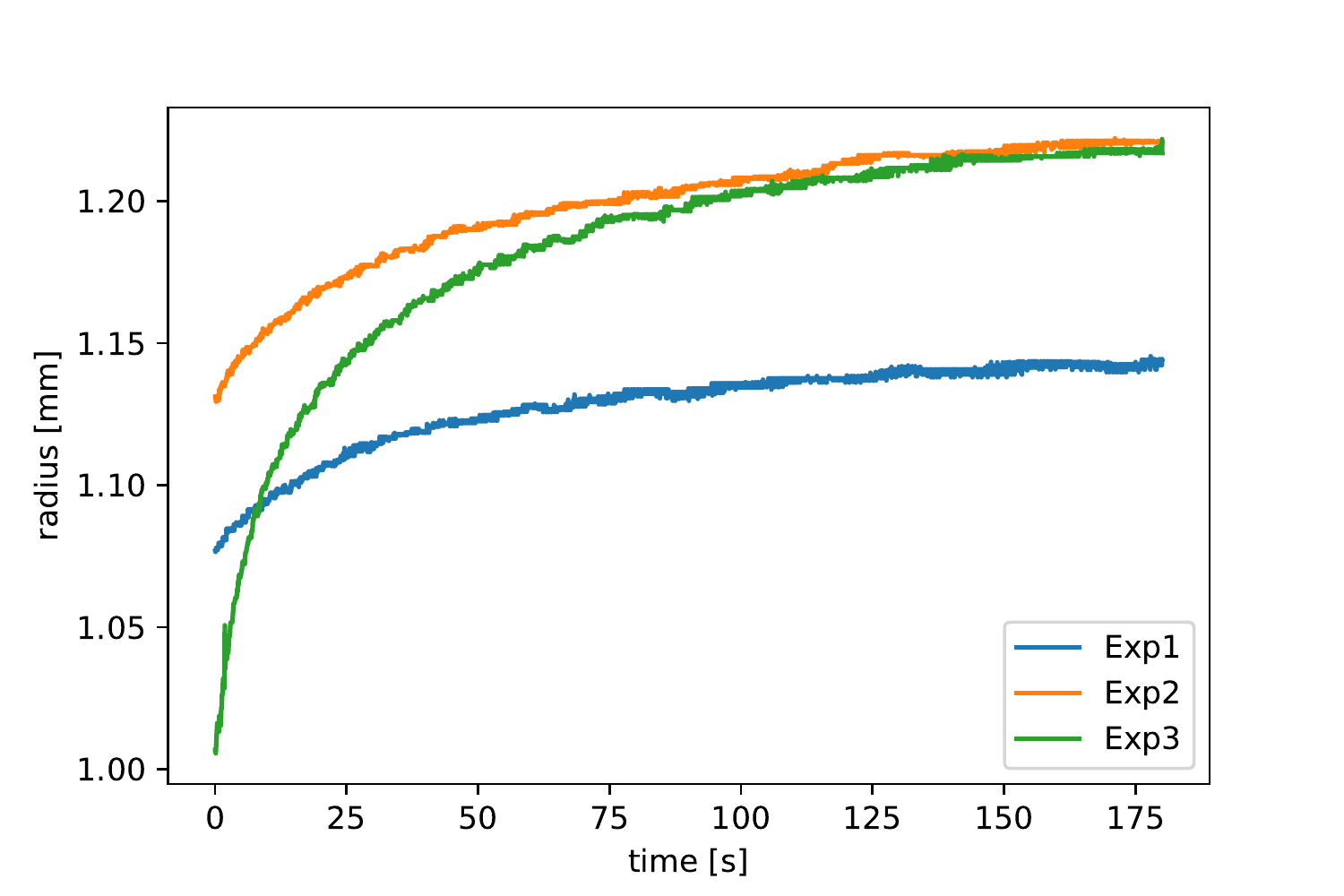}\label{fig:polymer-brush_80/radius}}
 \caption{Spreading of a water droplet on a polymer brush ($80\%$ humidity).}
\end{figure}

\paragraph{Spreading in a high humidity environment:} We first consider experimental data for a humidity of $80\%$ within the climate chamber. The experimental data for the volume reported in Fig.~\ref{fig:polymer-brush_80/volume} show that the droplet slowly loses volume mainly due to evaporation. Compared to the dynamics of the glycerol droplet on the silicon wafer in the previous section, the spreading on the polymer brush is extremely slow; see Fig.~\ref{fig:polymer-brush_80/radius}.\\

\begin{figure}[htb]
\subfigure[Empirical function.]{\includegraphics[width=6.5cm]{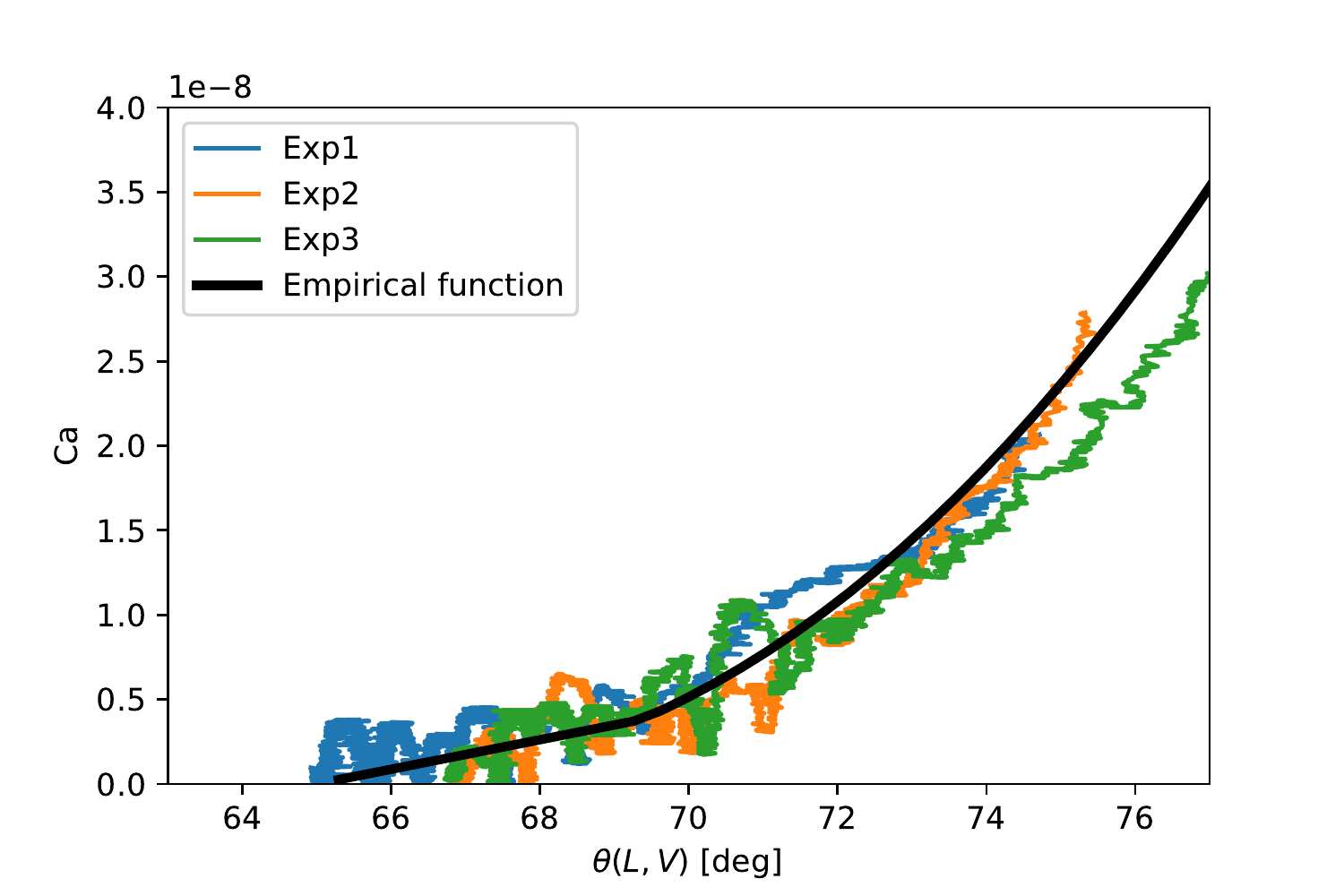}\label{fig:polymer-brush_80/individual_psi}}
\subfigure[ODE solution for the base radius.]{\includegraphics[width=6.5cm]{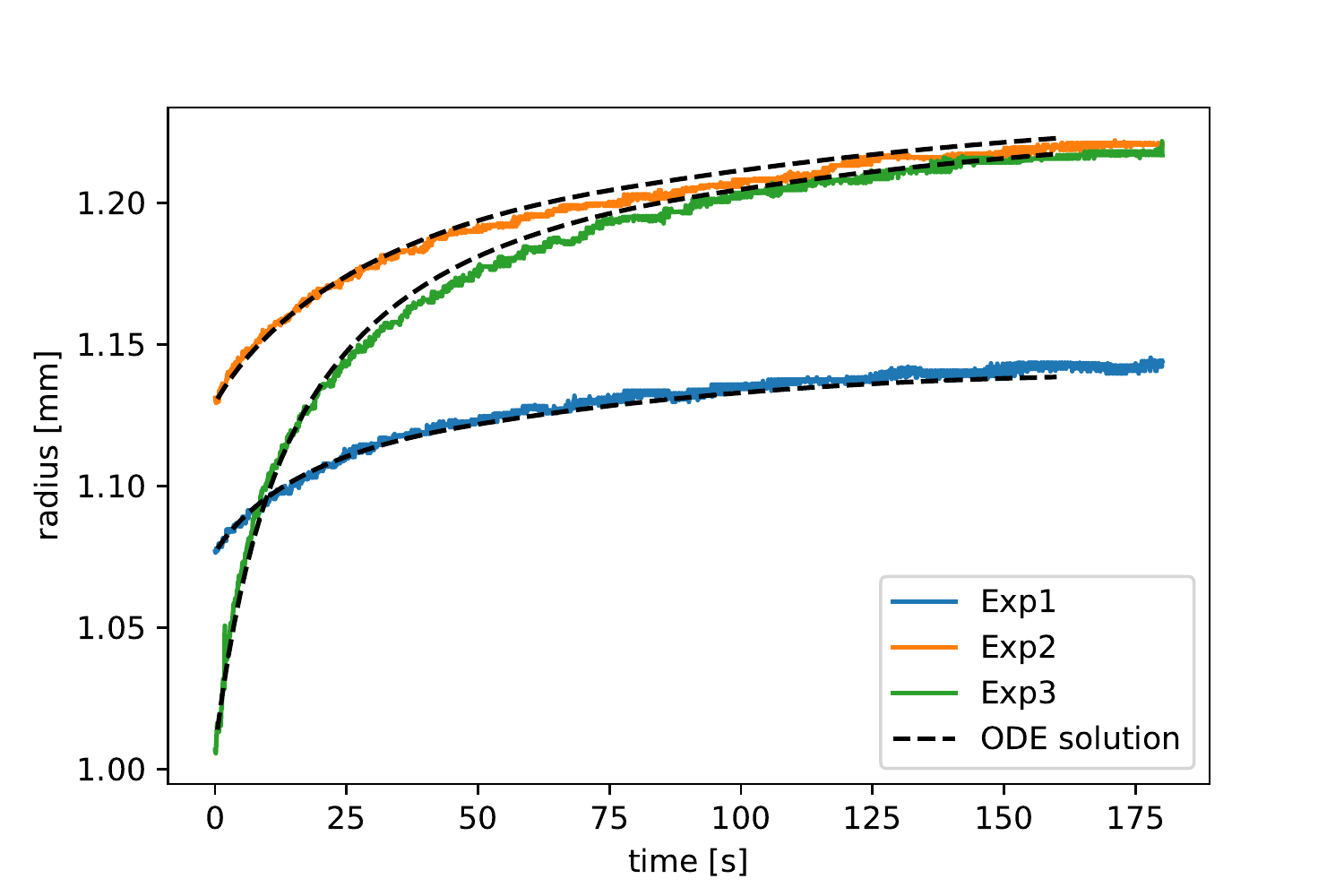}\label{fig:polymer-brush_80/solution_radius}}
\caption{Empirical function and ODE solution ($80\%$ humidity).}
\end{figure}

\begin{figure}[htb]
\centering
\includegraphics[width=8.0cm]{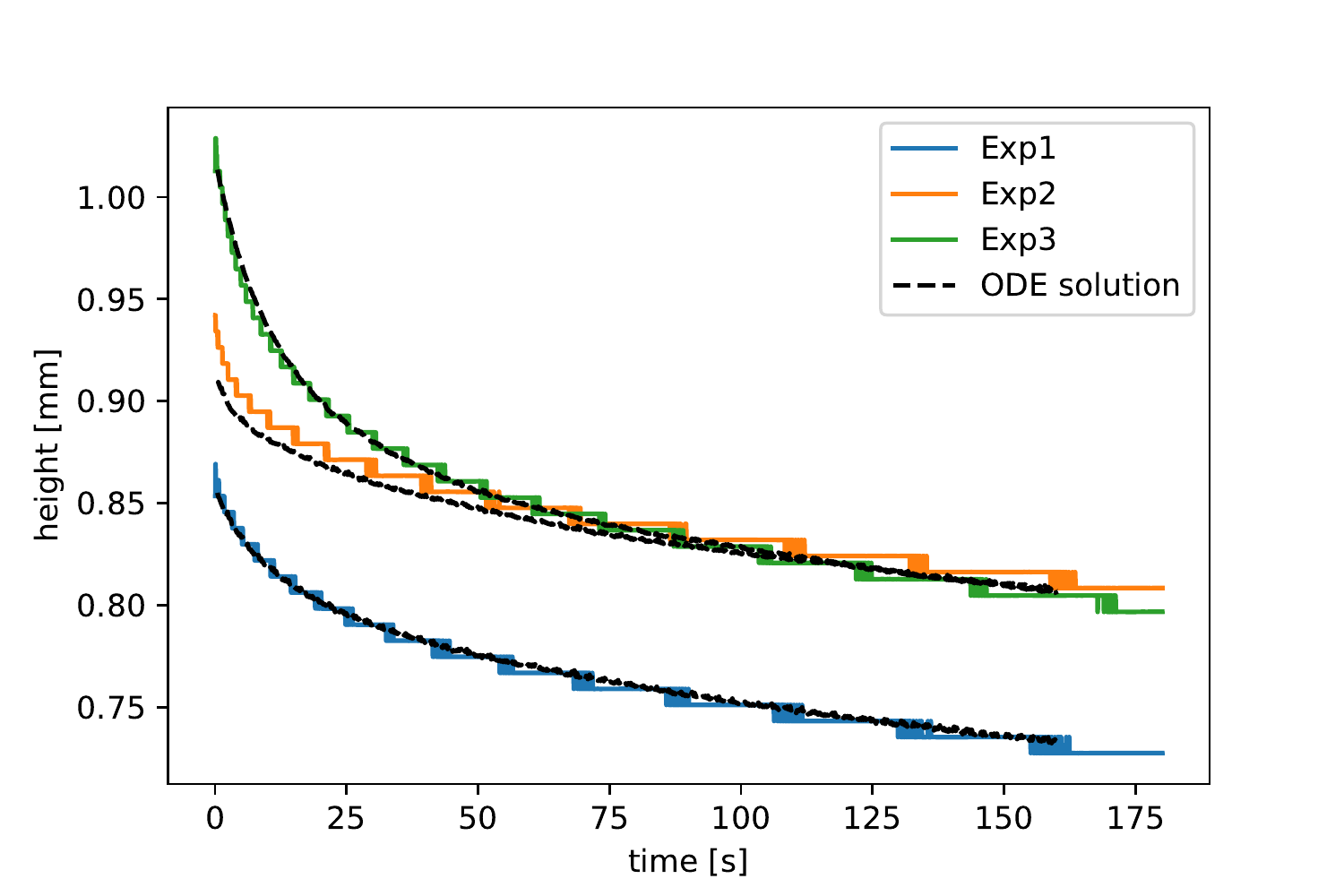}
\caption{Experimental data for the height compared to the ODE solution ($80\%$ humidity).}
\label{fig:polymer-brush_80/height}
\end{figure}

The experimental data for the capillary number as a function of the contact angle are shown in Figure~\ref{fig:polymer-brush_80/individual_psi}. It is found that the data collapse reasonably well onto a single curve which can be described by the empirical function
\begin{align}
\label{eqn:empirical_psi_polymer_brush_80}
\psi_{\text{emp}}(\theta) = \max\{1.1 \cdot 10^{-6} (\theta-\thetaeq)^{2.2}, \, 5 \cdot 10^{-8} (\theta-\thetaeq) \},
\end{align}
where $\thetaeq = 65^\circ$. This is remarkable since the physics of the interaction of the droplet with the substrate in the spreading process is expected to be quite complex. As can be seen from Figure~\ref{fig:polymer-brush_80/solution_radius}, the empirical relation \eqref{eqn:empirical_psi_polymer_brush_80} is able to describe the spreading dynamics for all considered repetitions of the experiment. Moreover, the predicted evolution of the droplet height agrees reasonably well with the experimental data; see Fig.~\ref{fig:polymer-brush_80/height}. Moreover, we find that the data for the polymer brush can be described by the same exponents as for the water-glycerol droplet on the bare silicon wafer.

\begin{figure}[htb]
\subfigure[Experimental data for the drop volume.]{\includegraphics[width=6.5cm]{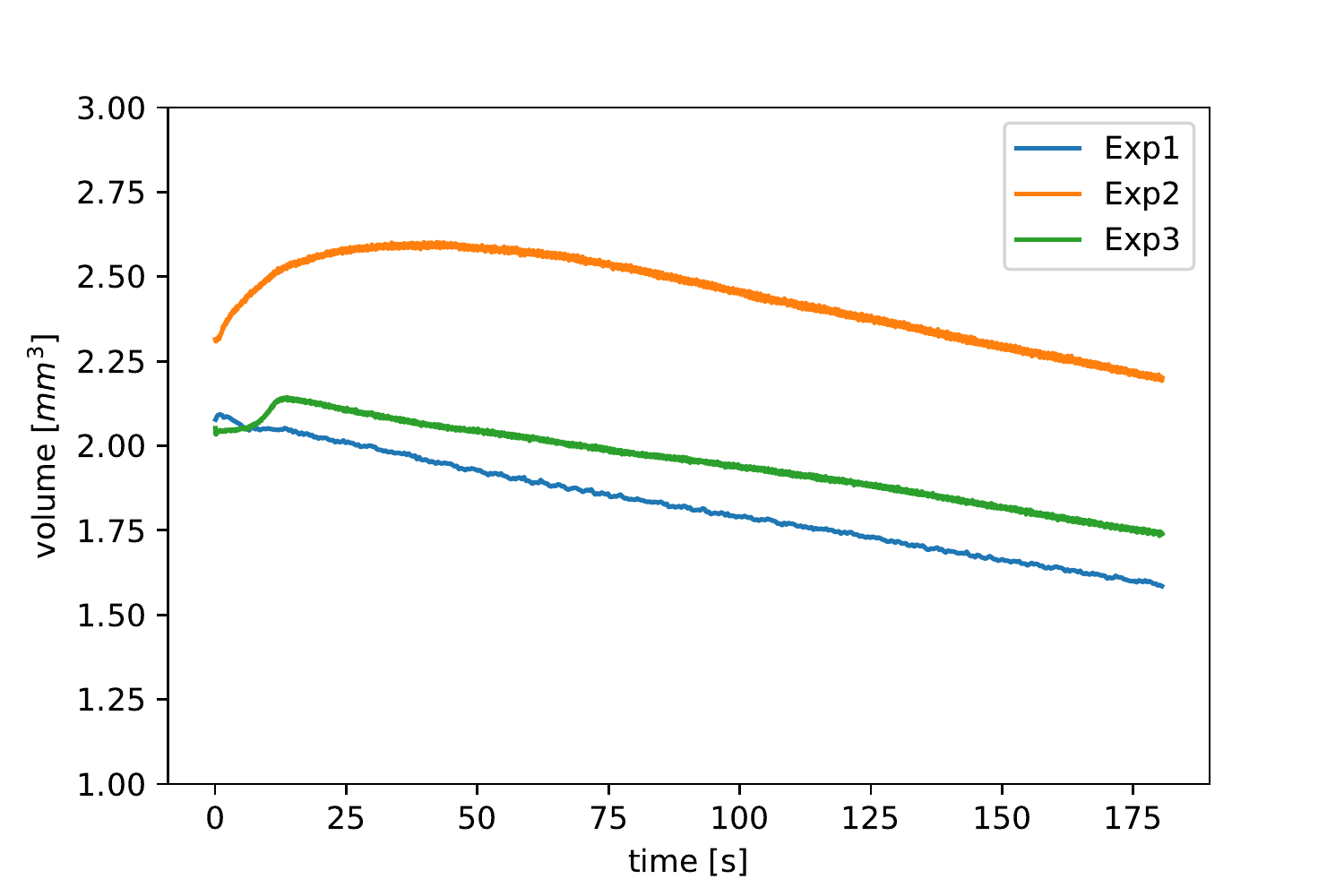}\label{fig:polymer-brush_50/volume}}
\subfigure[Experimental data for the drop base radius.]{\includegraphics[width=6.5cm]{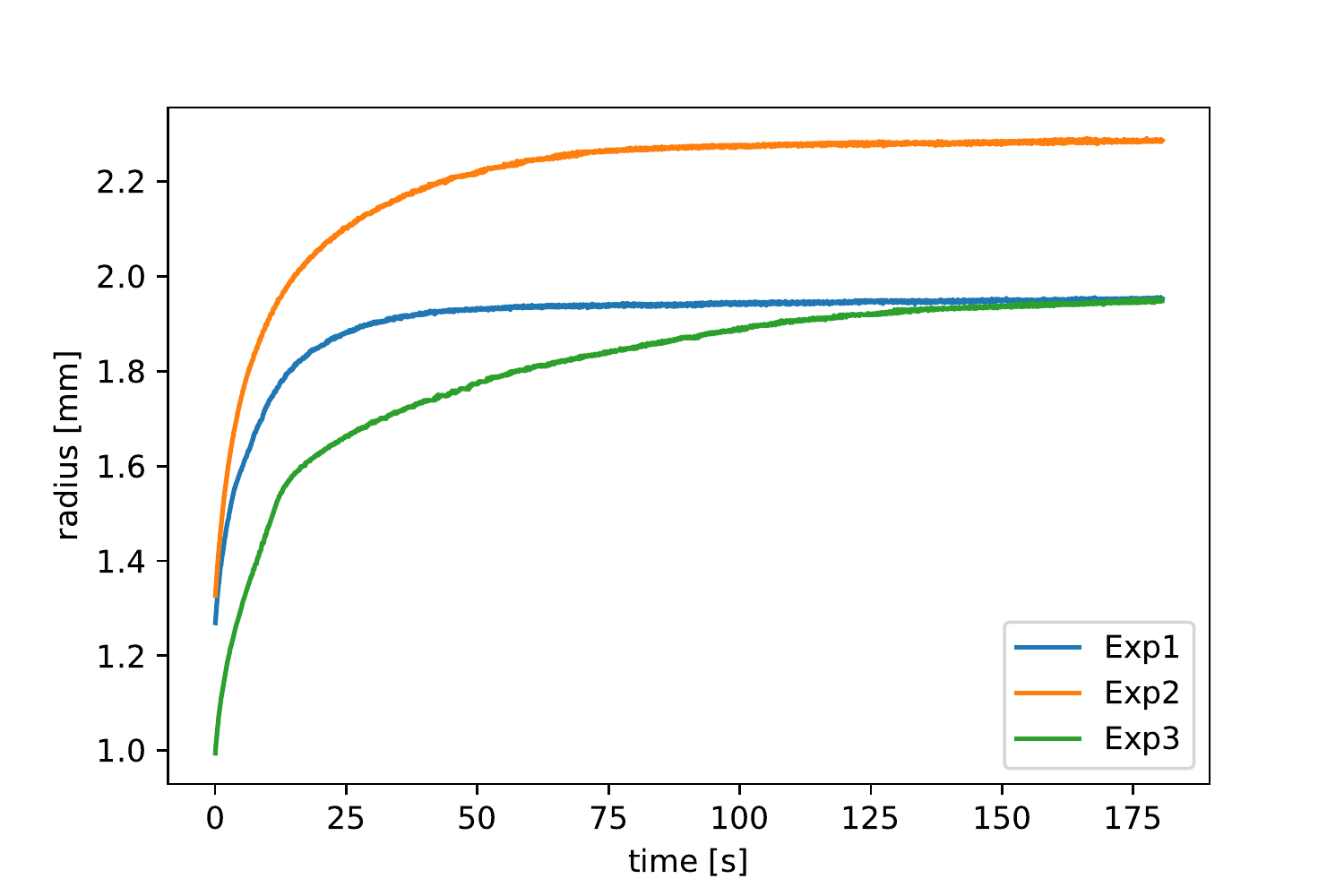}\label{fig:polymer-brush_50/radius}}
 \caption{Spreading of a water droplet on a polymer brush ($50\%$ humidity).}
\end{figure}

\begin{figure}[htb]
\subfigure[Experimental data for the drop height.]{\includegraphics[width=6.5cm]{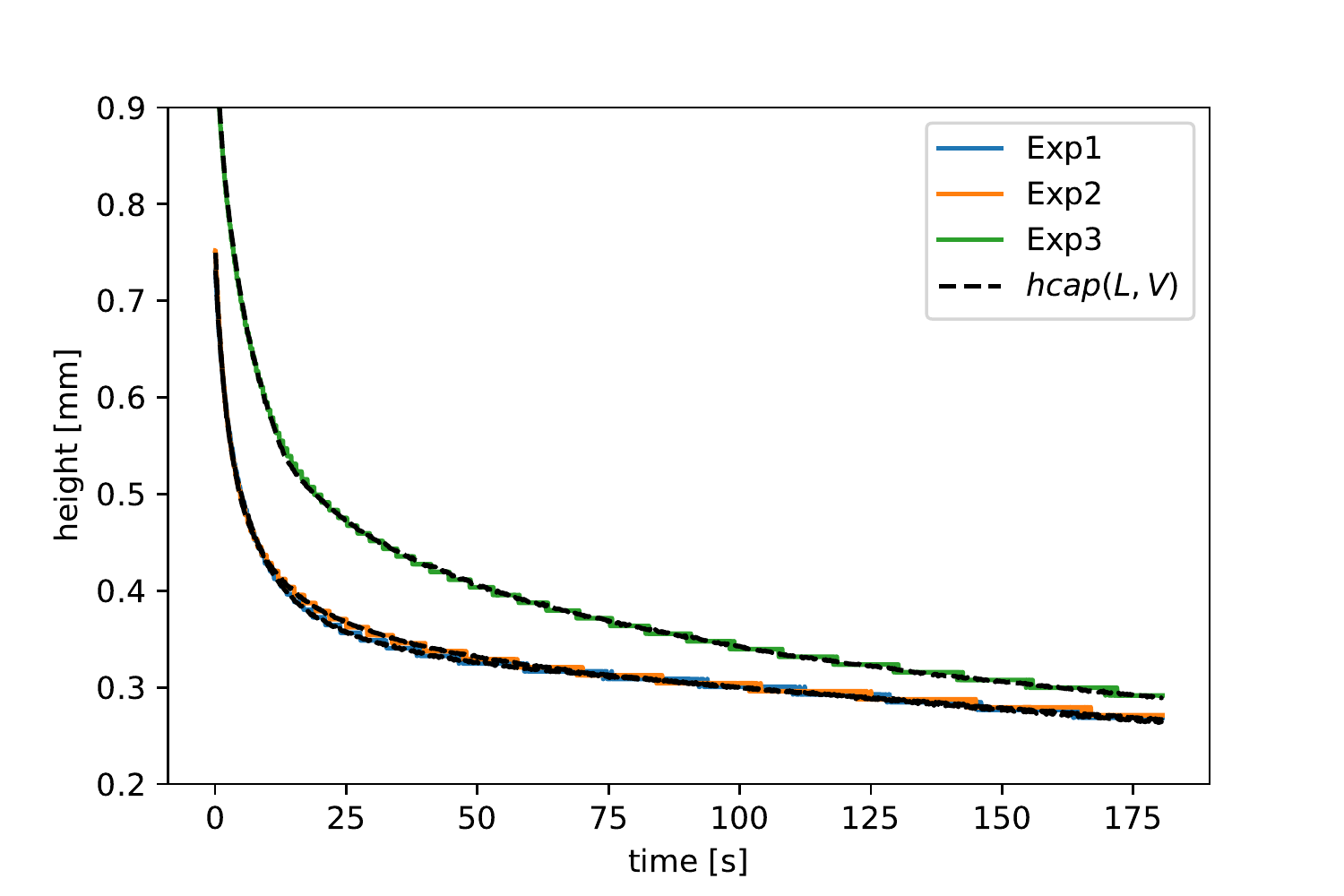}\label{fig:polymer-brush_50/height}}
\subfigure[Empirical function.]{\includegraphics[width=6.5cm]{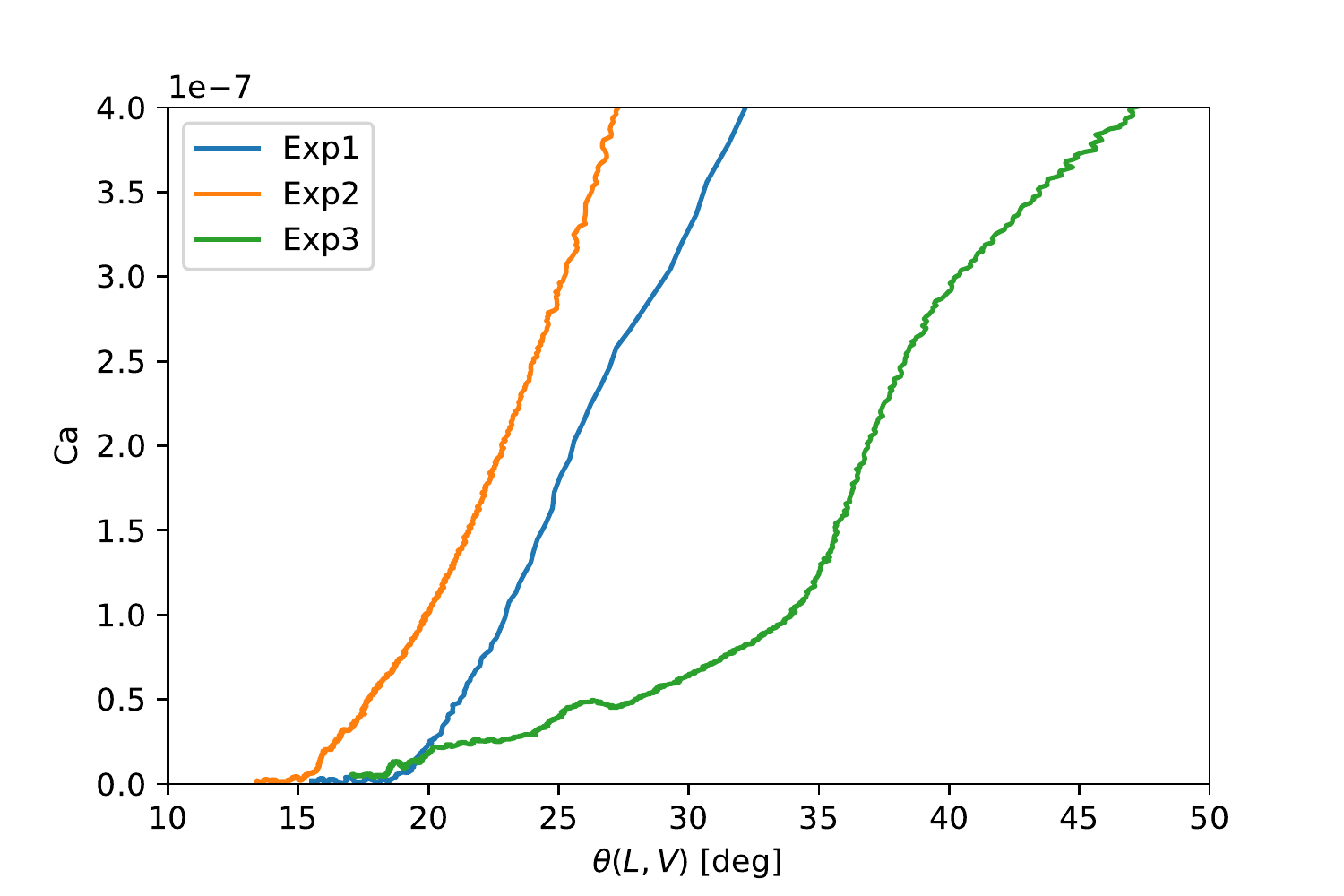}\label{fig:polymer-brush_50/individual_psi}}
 \caption{Droplet height and empirical function ($50\%$ humidity).}
\end{figure}

\paragraph{Spreading in a low humidity environment:} Finally, we consider the spreading in an environment with a reduced humidity of $50\%$. As can be expected, the evaporation rate is larger leading to a more rapid change in the drop volume; see Fig.~\ref{fig:polymer-brush_50/volume}. The experimental data for the drop height agree well with the assumption of a spherical cap geometry; see Fig.~\ref{fig:polymer-brush_50/height}. However, the experimental data for the relation between contact angle and capillary number in Fig.~\ref{fig:polymer-brush_50/individual_psi} clearly shows that \emph{no} universal relation of the form
\[ \Ca = \psi(\theta) \]
exists in this case. This is not surprising from a theoretical point of view since the presence of the droplet leads to an adaptation (i.e.\ swelling) of the polymer brush which in turn affects the wetting process. The latter effect can be expected to be more significant when the humidity is low since a high humidity leads to a pre-swelling of the polymer brush.

\section{Summary}
A generalization of the ODE model for the spreading of thin droplets from \cite{Gennes.1985} is introduced. The main assumption for the model to be applicable is a spherical cap shape of the droplet and a functional relation between the speed of the contact line and the macroscopic contact angle. The ability of the ODE model to describe the spreading kinetics of partially wetting liquids has been demonstrated for the viscous stage of spreading of a glycerol-water droplet on a silicon wafer and the spreading of a pure water droplet on a PNIPAm-brush coated silicon wafer. It is important to note that the ODE model can only \emph{predict} the spreading of spherical droplets in a range of parameters where a \emph{universal} relation between macroscopic contact angle and contact line speed exists.\\
\\
The assumption of a spherical cap geometry allows to compute the contact angle as a function of the base radius and the volume leading to a regularization of the data compared to a direct measurement of the contact angle.  The experimental data for the water-glycerol droplet show some variations in the relation between contact angle and contact line speed. It is found that the empirical relation cannot be described by a single exponent. Instead of that, there are at least two distinct stages with different exponents. Interestingly, there is a large deviation to the Cox-Voinov law and the empirical function by Kistler. The observed capillary number is one order of magnitude smaller than the prediction of the latter relations. This effect might be due to the polarity of the fluid or mixture effects and deserves further attention.\\
\\
Finally, the ODE model can also describe the spreading of water droplets on a PNIPAm-brush coated silicon wafer. Interestingly, we found a reasonable collapse of the data for the experiments in a high humidity environment. The exponents in the empirical function are found to be the same as for the water-glycerol droplet on a bare silicon wafer. Conversely, there is no universal empirical relation for the experiments in a reduced humidity environment. This clearly shows the need for a more detailed continuum mechanical modeling of the complex interaction between the dynamics of drop and the polymer brush.

\paragraph{Acknowledgments:} We kindly acknowledge the financial support by the German Research Foundation (DFG) within the Collaborative Research Center 1194 “Interaction of Transport and Wetting Processes” – Project-ID 265191195, subprojects A02, A05 and B01 and B02. Moreover, we thank T. Gambaryan-Roisman and M. Heinz (Project A04) for sharing their unpublished Droplet EvolutioN: Impact, Imbibition, Spreading and Evaporation (DENIISE) Matlab algorithm. We thank Longquan Chen and Elmar Bonaccurso for providing research data from \cite{Chen2014}.

\bibliography{ms.bbl}

\clearpage
\appendix

\section{Preparation of the polymer brush}
\label{sec:polymer_brush_preparation}
The polymer brush synthesis has been adapted from well-known protocols known in literature \cite{Ye2009}. The following outlines a brief summary of the preparation with respect to the used chemicals and preparation procedures.\\
\\
\textbf{Materials:} Copper(I)chloridee (99\%, sigma aldrich), allylamine (98\%, alfa aesar),  
triethylamine (TEA, 99\%, alfa aesar), chlorodimethylsilane (98\%, sigma aldrich), 
tris(2-dimethylaminoethyl)amine (ME\textsubscript{6}TREN, 99\%, alfa aesar), dimethylformamid (DMF, anhydrous 99.8\%, fischer scientific), toluene (anhydrous 99.85\%, acros organics) and Pt/C (10\% Pt, sigma adrich) were used as received unless specified otherwise. 
\textit{N}-Isopropylacrylamid (NIPAm, $98\%$) was purchased from Sigma-Aldrich and recrystallised (toluene/hexane 1:4)
before using. Surface-initiated polymerizations were performed from initiator modified silicon (P$/\,$bor \textless100\textgreater) 
substrates from Si-Mat that were coated with a native silicon oxide layer (wafer size: {24 x 24}$\,$\si{\milli\metre\squared}).\\
\\
\textbf{Methods:} Brush thicknesses were determined using a computer-controlled multiple angle null-ellipsometer 
(Accurion EP\textsuperscript{3} System) operating at \SI{658}{\nano\metre} between \SIrange{40}{68}{\degree}. Ellipsometric film thicknesses were 
calculated by a three-layer silicon/polymer brush/ambient model (refractive indices: $n_{\text{silicon}}=3.7, 
n_{\text{polymer}}=1.5$), assumed a isotropic and homogeneous polymer brush.\\
\\
\textbf{Polymer brush Synthesis:} The ATRP initiator 2-Bromo-2-methyl-\textit{N}-(3-[chloro(di-methyl)silyl]-propyl)propanamide (\textbf{1}) was 
obtained by coupling 2-bromo-2-methyl-\textit{N}-allylpropanamide\cite{Paripovic2011} and chlorodimethylsilane in 
presence of Pt/C (10\% Pt) under argon by refluxed overnight at \SI{50}{\celsius}.
\\
\\
Prior to the ATRP initiator immobilization on solid substrates, the Si wafer were cleaned 
using a Femto plasma system (Diener electronics, Germany) for a period of \SI{1}{\minute} at \SI{100}{\watt} after 
sonicated for \SI{30}{\minute} in aceton and ethanol and dried under nitrogen. The clean Si wafers were kept 
overnight in a $10\,$mM solution of 1 and TEA in anhydrous toluene. Subsequently the slides were extensively 
rinsed with chloroform and dried under nitrogen.
\\
\\
Subsequent to the immobilization of the initiatior Poly(\textit{N}-isopropylacrylamide) (PNIPAm) \cite{CMU-EDU} 
brushes were prepared according to literature procedure. \SI{37}{\milli\gram} (\SI{0,37}{\milli\mole}) copper(I)chloride,
\SI{94}{\micro\liter}
(\SI{0,35}{\milli\mole}) \ce{Me6TREN} and water (\SI{6}{\milli\liter}) were premixed in a Schlenk tube and degassed by 
three freeze-pump-thaw cycles. In another Schlenk tube, \SI{12}{\gram} (\SI{106}{\milli\mole}) NIPAm was mixed with DMF (\SI{24}{\milli\liter}) and degassed in the same way as the previous mixture after the ATRP-initiator-modified substrates were 
added under nitrogen. The catalyst solution of the first Schlenk tube was transferred to the monomer solution 
to initialize the polymerization. After \SI{8}{\hour} the substrates were removed from the polymerization reactor, 
rinsed with \SI{50}{\milli\liter} of the respective solvent and ethanol p.a. followed by soxhlet extraction (\SI{4}{\hour}, THF) 
and dried under a flow of nitrogen.

\section{Geometrical relations and exact drop shapes}
\label{sec:appendix_mathematics}
Let $L$ and $h$ be the base radius and the height of the spherical cap and $R$ be the radius of the corresponding sphere and $\theta$ be the contact angle. Then we have
\[ h = R(1-\cos \theta), \quad V = \frac{4}{3} \pi R^3 = \frac{\pi}{3} h^2 \underbrace{(3R-h)}_{\geq R}. \]
Since $L = R \sin \theta$, one can rewrite the equation for the height to find
\[ h = \frac{L(1-\cos \theta)}{\sin \theta} = L \tan(\theta/2). \]
Using the geometric relation \eqref{eqn:basic_geometric_relation}, the formula for the height of a spherical cap as a function of the volume and the contact angle follows:
\begin{align}
h_{\text{cap}}(\theta,V) = V^{1/3} g(\theta) \tan(\theta/2) =: V^{1/3} \tilde{g}(\theta).
\end{align}
Note that the function $\tilde{g}$ is monotonically increasing on $[0,\pi]$ with $\tilde{g}(0) = 0$ and $\tilde{g}(\pi) = (6/\pi)^{1/3}$. An equivalent formulation in terms of $L$ and $V$ is obtained from \eqref{eqn:basic_geometric_relation} according to
\begin{align}
h_{\text{cap}}(L,V) = V^{1/3} \tilde{g}\left( g^{-1}\left(\frac{L}{V^{1/3}}\right)\right) = L \tan\left(\frac{1}{2} \, g^{-1}\left(\frac{L}{V^{1/3}}\right)\right).
\end{align}

\subsection{Exact drop shapes in stationary state}
To quantify the influence of gravity and thereby justify the assumption of a spherical cap shaped droplet, the exact drop shapes are computed based on a continuum mechanical model. Using height functions for the representation of the liquid-gas interface yields the following boundary value problem for the interface shape $h(r)$:
\begin{align}
h''
&=
(1+h'^2)^{3/2} \left(\frac{\rho g}{\sigma} h- \frac{p_0}{\sigma} - \frac{h'}{r\sqrt{1+h'^2}}\right) \qquad \text{for } r \,\in [-R,R], \label{eq:drop_bvp}\\
 h(-R) &= 0, \qquad h'(-R) = \tan \theta.  \label{eq:drop_bvp_bc}
\end{align}
The solution also has to satisfy the integral condition
\begin{align}\label{eq:drop_volume}
  V &= \int_0^R \int_0^{2\pi} h \, \mathrm{d}\varphi\, r\mathrm{d}r,
\end{align}
where $V$ is the volume of the drop and $R$ is the radius of the drop which is unknown a priori. Note that the integral condition makes the above boundary value problem an integro-differential equation, while the size of the interval is also part of the solution. Furthermore, the computation of the drop shape comprises an inverse problem with the parameter $p_0$, which is the pressure inside the drop.\newline
The inverse problem described above is solved using an algorithm from \cite{Gruending2020} that is based on the method of nested intervals for $p_0$ to obtain the interface shape $h$ and the drop pressure $p_0$. Starting from an initial pressure guess. with this approach the interface shape is subsequently improved from an initial interface shape of a spherical cap. Overall, the algorithm allows to compute the full interface shape from the contact angle $\theta$, the drop volume $V$, and the corresponding Bond number $\text{Bo}$. For a more detailed description see \cite{Gruending2020}.\newline
The described algorithm has been used to compute droplet shapes for varying Bond numbers and contact angles. Figure \ref{fig:drop_shapes} shows the dimensionless drop heights over the Bond number for various contact angles. For each Bond number and contact angle, one inverse problem has been solved as described above. Here, the length scale which is used for the Bond number in \eqref{eq:bond_number} is the radius of a volume-equivalent drop with a contact angle of \SI{90}{\degree}, i.e. $l := \sqrt[3]{3V/2\pi}$. The drop height is scaled, using the height of the corresponding spherical cap shaped drop, i.e. the scaled drop height is given by $h^\star = h/h_{\text{cap}}(L,V)$.

\end{document}